\documentclass[iop]{emulateapj}

\usepackage{xcolor}

\usepackage{apjfonts}
\usepackage{graphicx}
\usepackage{epstopdf}
\usepackage{ulem}
\usepackage[utf8]{inputenc}
\usepackage[T1]{fontenc}
\usepackage{url}
\usepackage{hyperref}
\usepackage{mdwlist}
\usepackage{amsmath} 
\usepackage{enumitem}
\usepackage{lineno}

\makeatletter
\def\url@leostyle{%
 \@ifundefined{selectfont}{\def\UrlFont{\sf}}{\def\UrlFont{\small\ttfamily}}}
\makeatother
\begin{document}

\newcommand{\ls}{{_<\atop^{\sim}}}
\newcommand{\gs}{{_>\atop^{\sim}}}
\def \spose#1{\hbox  to 0pt{#1\hss}}  
\def \ls{\mathrel{\spose{\lower 3pt\hbox{$\sim$}}\raise  2.0pt\hbox{$<$}}}
\def \gs{\mathrel{\spose{\lower  3pt\hbox{$\sim$}}\raise 2.0pt\hbox{$>$}}}
\newcommand{\Ha}{\hbox{{\rm H}$\alpha$}}
\newcommand{\Hb}{\hbox{{\rm H}$\beta$}}
\newcommand{\Ovi}{\hbox{{\rm O}\kern 0.1em{\sc vi}}}
\newcommand{\OIII}{\hbox{[{\rm O}\kern 0.1em{\sc iii}]}}
\newcommand{\OII}{\hbox{[{\rm O}\kern 0.1em{\sc ii}]}}
\newcommand{\NII}{\hbox{[{\rm N}\kern 0.1em{\sc ii}]}}
\newcommand{\SII}{\hbox{[{\rm S}\kern 0.1em{\sc ii}]}}
\newcommand{\angstrom}{\textup{\AA}}
\newcommand\ionn[2]{#1$\;${\scshape{#2}}}

\font\btt=rm-lmtk10


\title{Evidence of wind signatures in the gas velocity profiles of Red Geysers }

\shorttitle{AGN-driven winds in red geysers}

\shortauthors{Roy et al.}


\author{Namrata Roy\altaffilmark{1}\dag, 
Kevin Bundy\altaffilmark{1,2},
Rebecca Nevin\altaffilmark{3},
Francesco Belfiore\altaffilmark{4},
Renbin Yan\altaffilmark{5},
Stephanie Campbell\altaffilmark{6},
Rogemar A. Riffel\altaffilmark{7,8}, 
Rogerio Riffel\altaffilmark{9,8}, 
Matthew Bershady\altaffilmark{10,11,12}, 
Kyle Westfall\altaffilmark{2}, 
Niv Drory\altaffilmark{13},
Kai Zhang\altaffilmark{14}}

\altaffiltext{1} {Department of Astronomy and Astrophysics, University of California, 1156 High Street, Santa Cruz, CA 95064, USA.}\altaffiltext{\dag}{naroy@ucsc.edu}
\altaffiltext{2}{UCO/Lick Observatory, Department of Astronomy and Astrophysics, University of California, 1156 High Street, Santa Cruz, CA 95064, USA.}
\altaffiltext{3} {Center For Astrophysics, Harvard $
\&$ Smithsonian, 60 Garden Street, Cambridge, MA 02138, USA.}
\altaffiltext{4} {INAF – Osservatorio Astrofisico di Arcetri, Largo E. Fermi 5,
I-50157, Firenze, Italy.}
\altaffiltext{5}{Department of Physics and Astronomy, University of
Kentucky, 505 Rose St., Lexington, KY 40506-0055, USA.}
\altaffiltext{6}{School of Physics and Astronomy, University of St Andrews, North Haugh, St Andrews, Fife KY16 9SS, UK.}
\altaffiltext{7}{Departamento de F\'isica, CCNE, Universidade Federal de Santa Maria, 97105-900, Santa Maria, RS, Brazil.}
\altaffiltext{8}{Laborat\'orio Interinstitucional de e-Astronomia - LIneA, Rua Gal. Jos\'e Cristino 77, Rio de Janeiro, RJ - 20921-400, Brazil.}
\altaffiltext{9}{Departamento de Astronomia, Instituto de F\'\i sica, Universidade Federal do Rio Grande do Sul, CP 15051, 91501-970, Porto Alegre, RS, Brazil. }
\altaffiltext{10}{University of Wisconsin - Madison, Department of Astronomy,
475 N. Charter Street, Madison, WI 53706-1582, USA.}
\altaffiltext{11}{South African Astronomical Observatory, PO Box 9, Observatory 7935, Cape Town, South Africa.}
\altaffiltext{12}{Department of Astronomy, University of Cape Town, Private Bag X3, Rondebosch 7701, South Africa.}
\altaffiltext{13}{McDonald Observatory, The University of Texas at Austin,
2515 Speedway, Stop C1402, Austin, TX 78712, USA.}
\altaffiltext{14}{Lawrence Berkeley National Laboratory, 1 Cyclotron Road, Berkeley, CA 94720, USA.}

\begin{abstract}

Spatially resolved spectroscopy from SDSS-IV MaNGA survey has revealed a class of quiescent, relatively common early-type galaxies, termed ``red geysers'', that possibly host large scale active galactic nuclei driven winds. Given their potential importance in  maintaining low level of star formation at late times, additional evidence confirming that winds are responsible for the red geyser phenomenon is critical. In this work, we present follow-up observations with the Echellette Spectrograph and Imager (ESI) at the Keck telescope of two red geysers (z$<$0.1) using multiple long slit positions to sample different regions of each galaxy. Our ESI data with a spectral resolution (R) $\sim$ 8000 improves upon MaNGA's resolution by a factor of four, allowing us to resolve the ionized gas velocity profiles along the putative wind cone with an instrumental resolution of $\rm \sigma = 16~km~s^{-1}$. The line profiles of H$\alpha$ and [NII]$\rm \lambda 6584$ show asymmetric shapes that depend systematically on location $-$ extended blue wings on the red-shifted side of the galaxy and red wings on the opposite side. We construct a simple wind model and show that our results are consistent with geometric projections through an outflowing conical wind oriented at an angle towards the line of sight. An alternative hypothesis that assigns the asymmetric pattern to ``beam-smearing'' of a rotating, ionized gas disk does a poor job matching the line asymmetry profiles. While our study features just two sources, it lends further support to the notion that red geysers are the result of galaxy-scale winds.

\end{abstract}



\section{Introduction} \label{sec:Introduction}

``Maintenance'' or ``radio'' mode feedback resulting from low to moderate luminosity active galactic nuclei (AGN) has been proposed as a means for maintaining low levels of star formation (log SFR $< \rm -2~M_{\odot}~yr^{-1}$) at late times, thereby explaining the massive increase in the number of red galaxies since $\rm z\sim2 $ \citep{croton06, bell04, bundy06, faber07, ilbert10, moustakas13}. These radio AGNs are thought to be radiatively inefficient, accreting at low rates and depositing most of their energy to the surroundings via momentum driven winds or radio jets \citep{binney95, ciotti01, croton06, bower06, ciotti07, ciotti10, mcnamara07, cattaneo09, fabian12, yuan14, heckman14}. This input energy heats ambient gas that might otherwise cool and form stars, thus maintaining quenched galaxies in a passive state. Although evidence for radio mode AGN feedback has been observed in the form of large bubbles of ionized gas and radio jets in the centers of massive clusters and giant radio galaxies \citep{cattaneo09, dunn06,  fabian94, fabian12, fabian06, mcnamara07}, it has been difficult to find evidence for this mechanism operating in lower mass halos that host typical quiescent galaxies (halo mass < $\rm 10^{12}~M_{\odot}$).


We have been studying a population of moderate mass (log $\rm M_{\star}/M_{\odot} \sim 10.5$) red quiescent galaxies ($NUV-r > 5$), known as ``red geysers'' \citep{cheung16, roy18} that may hold clues in this regard. Identified in low redshift integral field spectroscopy from the Sloan Digital Sky Survey-IV (SDSS-IV) Mapping Nearby Galaxies at Apache Point Observatory (MaNGA) survey \citep{bundy15}, these galaxies are characterized by bisymmetric equivalent width (EW) maps of strong emission lines like H$\alpha$, [NII]$\rm\lambda\lambda$6548,84 and [OIII]$\rm\lambda\lambda$4959,5007 which appear to be tracing large scale ionized gas outflows. This distinctive emission pattern roughly aligns with the gas kinematic axis but is strongly misaligned with the stellar velocity gradient. The gas velocity values reach $\sim$300$~\rm km~s^{-1}$ compared to less than 40$\rm ~ km~s^{-1}$ in the stars.
These galaxies lack star formation with average log SFR [$\rm M_{\odot} / yr ] < -2$ \citep[using GALEX+SDSS+WISE,][]{salim16}. They additionally show a high value of the 4000-\AA~break index (D4000) with an average value > 2.0, thus providing further evidence that young stars are absent in the galaxy. Spatially resolved Baldwin-Phillips-Terlevich diagrams \citep[BPT, ][]{baldwin} indicate wide-spread ionization with line ratios consistent with a combination of LINERs (Low Ionization Nuclear Emission Region)  and Seyfert values. The morphology of these galaxies are spheroidal with a high sersic index (n $\geq$~3). 

Although the observed characteristics of the red geysers can be explained by a centrally driven wind, early type galaxies with accreted gas disks \citep{chen16, sarzi06, davis13, bryant19, lagos14, lagos15, starkenburg19,duckworth20} can also show similar kinematic misalignment and emission features due to rotation of the gaseous material in the disk. 
However, 95\% of the red geyser sample are fast rotator early-type galaxies from the \cite{graham18} catalog \citep{roy18} and it is considerably difficult for a gas disk to be in equilibrium if it is misaligned with the stellar kinematic axis because of the axisymmetric nature of the fast rotators \citep[although see ][]{vandevoort, davis16}. The red geysers are also selected to have axis ratio b/a > 0.4 \citep{roy18} with no visible dust lanes, as seen from SDSS imaging, in order to exclude any edge-on disks in the sample. A disturbed disk with a chaotic accretion scenario is still possible and further investigation is therefore needed to confirm otherwise. 



In \cite{roy18} we reported evidence for the presence of faint radio AGNs in the red geyser galaxies. Stacked 1.4 GHz radio flux from very large array (VLA) Faint Images of the Radio Sky at Twenty-Centimeters (FIRST) survey shows significantly ($>5\sigma$) higher radio flux in red geysers than a matched control sample. Red geysers also show a three times higher radio detection rate than the control. 
 \cite{roy18} shows that this radio emission indicates low-luminosity radio AGNs ($\rm L_{1.4 GHz} \sim 10^{22} - 10^{23}~W/Hz$) with radiatively inefficient accretion (Eddington scaled accretion rate $\lambda \sim 10^{-4}$). Recently \cite{duckworth20} has indicated a tentative correlation between enhanced AGN activity and misaligned gas disks in low mass galaxies ($\rm log~M~< 10.2~M_\odot$) using IllustrisTNG simulations. However no such trend has been found in high mass quenched population having similar mass as the red geyser sample. Hence the enhancement in radio-AGN activity seen in the red geysers cannot be immediately attributed to the phenomena of misaligned gas disks. \cite{riffel19} studied the launching of these proposed winds with Gemini GMOS (Gemini Multi Object Spectrograph) observations of the prototypical red geyser in \cite{cheung16} to constrain the gas kinematics in the nuclear region. They observed the emission line flux distributions and gas kinematics within the inner 1$''$ to be distinct and misaligned from that of the outer regions, $5''$ away from center, a result that may indicate precession of the accretion disk.
 


\cite{cheung16} presented a variety of evidence including dynamical modeling and geometric arguments that lend support for an interpretation of red geysers as an AGN-driven wind phenomenon \citep[see also ][]{gomes16}. However, it is important to seek out additional lines of evidence to distinguish outflows from rotation.
In this work, we examine further evidence for the wind interpretation through followup Keck spectroscopy of red geyser galaxies with higher spectral resolution (R$\sim8000$) than the original MaNGA data (R$\sim 2000$). The improved resolution allows us to search for detailed kinematic signatures of outflowing winds which are blurred at MaNGA's instrumental resolution. We detect asymmetric emission lines that vary in a systematic manner along the kinematic major axis which likely indicate the specific geometry of the wind along the line of sight. 
For the two red geysers (MaNGA ID: 1-217022 and 1-145922) studied here with the Keck Echellette Spectrograph and Imager (ESI) instrument, we find results consistent with our simple wind model. An alternative hypothesis that assigns the asymmetric pattern to “beam-smearing” of a rotating, ionized gas disk is not favored by our data.

Throughout this paper, we assume a flat cosmological model with $H_{0} = 70$ km s$^{-1}$ Mpc$^{-1}$, $\Omega_{m} = 0.30$, and  $\Omega_{\Lambda} =0.70$, and all magnitudes are given in the AB magnitude system. 


\section{Observation and Data acquisition} \label{sec:data}



\subsection { The MaNGA survey}

We use observations from the SDSS-IV MaNGA survey \citep{bundy15, drory15, law15, yan16b, sdss16, blanton17}. MaNGA is an integral field spectroscopic survey that provides spatially resolved spectroscopy for nearby galaxies ($\rm z\sim0.03$) with an effective spatial resolution of $2.4''$ (full width at half-maximum; FWHM). The MaNGA survey uses the SDSS 2.5 meter telescope in spectroscopic mode \citep{gunn06} and the two dual-channel BOSS spectrographs \citep{smee13} that provide continuous wavelength coverage from the near-UV to the near-IR: $\rm3,600-10,000$ \AA. The spectral resolution varies from $\rm R\sim1400$ at 4000~\AA~ to $\rm R\sim2600$ at 9000~\AA. An $r$-band signal-to-noise $(SN)$ of $\rm 4-8$~\AA$^{-1}$ is achieved in the outskirts (i.e., $\rm1-2~R_{e}$) of target galaxies with an integration time of approximately 3-hr. 
MaNGA has observed roughly 10,000 galaxies with $\rm \log~(M_*/ M_{\odot})\gs9$ across $\sim$ 2700 deg$^{2}$ over its 6~yr duration. In order to balance radial coverge versus spatial resolution, MaNGA observes two thirds of its galaxy sample to $\sim$ 1.5~R$_e$ and one third to 2.5~R$_e$. The MaNGA target selection is described in detail in \cite{wake17}.

The raw data are processed with the MaNGA Data Reduction Pipeline \citep[DRP,][]{law16, yan16}. An individual row-by-row algorithm is used to extract the fiber flux and derive inverse variance spectra from each exposure, which are then wavelength calibrated, flat-fielded and sky subtracted. We use the MaNGA sample and data products drawn from the MaNGA Product Launch-8 \citep[MPL-8, cf. Table 1 from][]{law2020a}. We use spectral measurements and other analyses carried out by version 2.3.0 of the MaNGA Data Analysis Pipeline (DAP). 
The data we use here are based on DAP analysis of each spaxel in the MaNGA datacubes. The final output from the DAP are gas and stellar kinematics, emission line properties and stellar absorption indices. All the spatially resolved 2D maps shown in the paper are outputs from the DAP. An overview of the DAP used for DR15 and its products is described by \cite{westfall19}, and assessments of its emission-line fitting approach is described by \cite{belfiore19}.

We use ancillary data drawn from the NASA-Sloan Atlas\footnote{\href{http://www.nsatlas.org}{http://www.nsatlas.org}} (NSA) catalog which reanalyzes images and derives morphological parameters for local galaxies observed in Sloan Digital Sky Survey imaging \citep{blanton11}. It compiles spectroscopic redshifts, UV photometry (from GALEX; \citealt{martin05}), stellar masses \citep{blanton07}, and structural parameters. 

\subsection {Keck ESI data}

ESI is a visible-wavelength, faint-object, imager and single-slit spectrograph in operation at the Cassegrain focus of the Keck II telescope since 1999. ESI has three modes of observations: echellete mode, low dispersion mode and direct imaging mode. The mode of operation used here is the echellette mode which provides cross-dispersed spectroscopy mode at a resolving power of up to $\rm R=13,000$. Slits are 20 arcseconds in length and are available with widths of 0.3, 0.5, 0.75, 1.0, 1.25 and 6.0 arcseconds with varying spatial scales and velocity resolution.The echellete mode disperses the light into ten orders with a dispersion ranging from $\sim$ 0.30 \AA/pixel in order six (red) to $\sim$ 0.16 \AA/pixel in order 15 (blue), while maintaining a roughly constant dispersion of 11.5 $\rm km s^{-1} pixel^{-1}$ in velocity across all orders. The orders are curved due to the distortion within the prisms. ESI has a wide spectral coverage spanning from 0.39 to 1.1 micrometer in a single exposure and a velocity resolution as low as 22 $\rm km~s^{-1}$ FWHM (using the 0.3 arcsec wide echellette slit). An Epps refracting camera and a single 2K$\times$4K detector are used for all three modes.

Utilizing two half nights in February 2017, we followed up 2 red geysers from our MaNGA sample. These targets completely fill ESI's 20$''$ slit. We were able to observe both targets with multiple slit positions 
to map out different parts of the galaxy. For the first target galaxy with ID 1-217022, which is the prototypical red geyser ``Akira'' from \cite{cheung16}, we observed 3 slit positions while for the second target (ID 1-145922), we observed 2 slit positions. For each slit position, we integrated for
1.5$-$2 hours (in source and sky combined), nodding to sky positions every 5 minutes to enable quality sky subtraction. However, for the second slit for the second target, a combination of high clouds and a lower integration time resulted in an unacceptably low signal-to-noise spectra. Hence, we discarded that slit for our present analyses. We took measurements of the standard star G191b2b at the beginning of each night with the same instrumental setting with an exposure of three minutes to perform flux calibration. The width of the slit for all observations was 0.5$''$ yielding a velocity resolution $\rm \sim~37~km~s^{-1}$ or R $\sim$ 8000.  The median seeing FWHM over the course of the observations was 0.9$''$.
This should be compared to MaNGA's R $\sim$ 2000 spectral resolution and its effective spatial
resolution of 2.4$''$. 

\begin{figure}[h!!!] 
\centering
\graphicspath{{./plots/}}
\includegraphics[width = 0.5\textwidth]{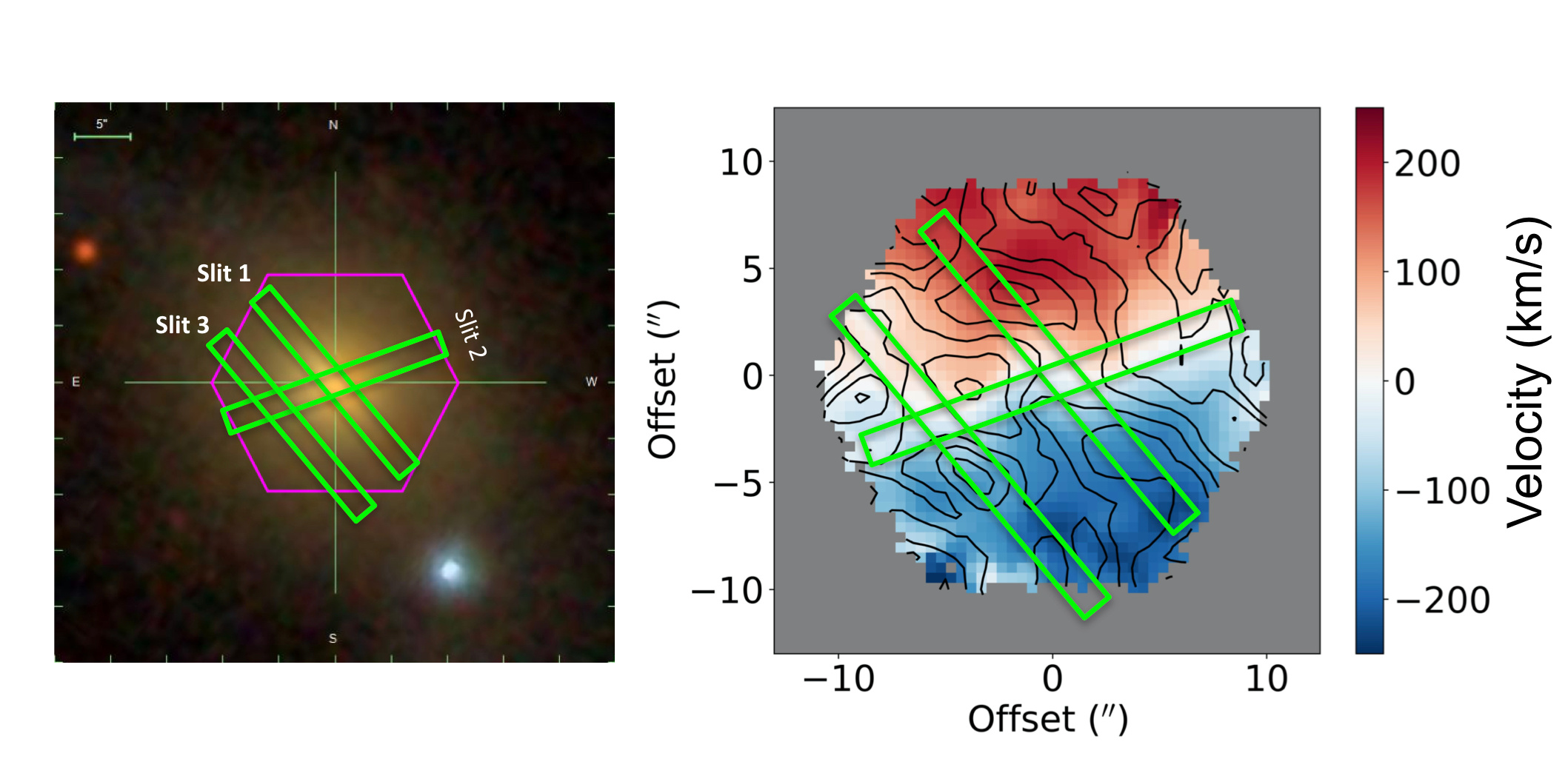}
\caption{  The Keck ESI slit positions for the first target galaxy (MaNGA-ID: 1-217022). The left and the right panels show the three slit positions overlaid on the SDSS optical image and the MaNGA ionized gas velocity map respectively. The black contours in the velocity map shows the H$\alpha$ equivalent width (EW). Slit 1, placed at an angle of 40$^{\circ}$ from North to East, traces the bi-symmetric emission feature. Slit 2 is placed at an angle of 110$^{\circ}$ from North to East, sampling the central parts of the galaxy. Slit 3 is placed parallel to slit 1 with an offset of $\sim\rm4.9''$.
\label{fig:slit1}}
\end{figure}

\begin{figure}[h!!!] 
\centering
\graphicspath{{./plots/}}
\includegraphics[width = 0.5\textwidth]{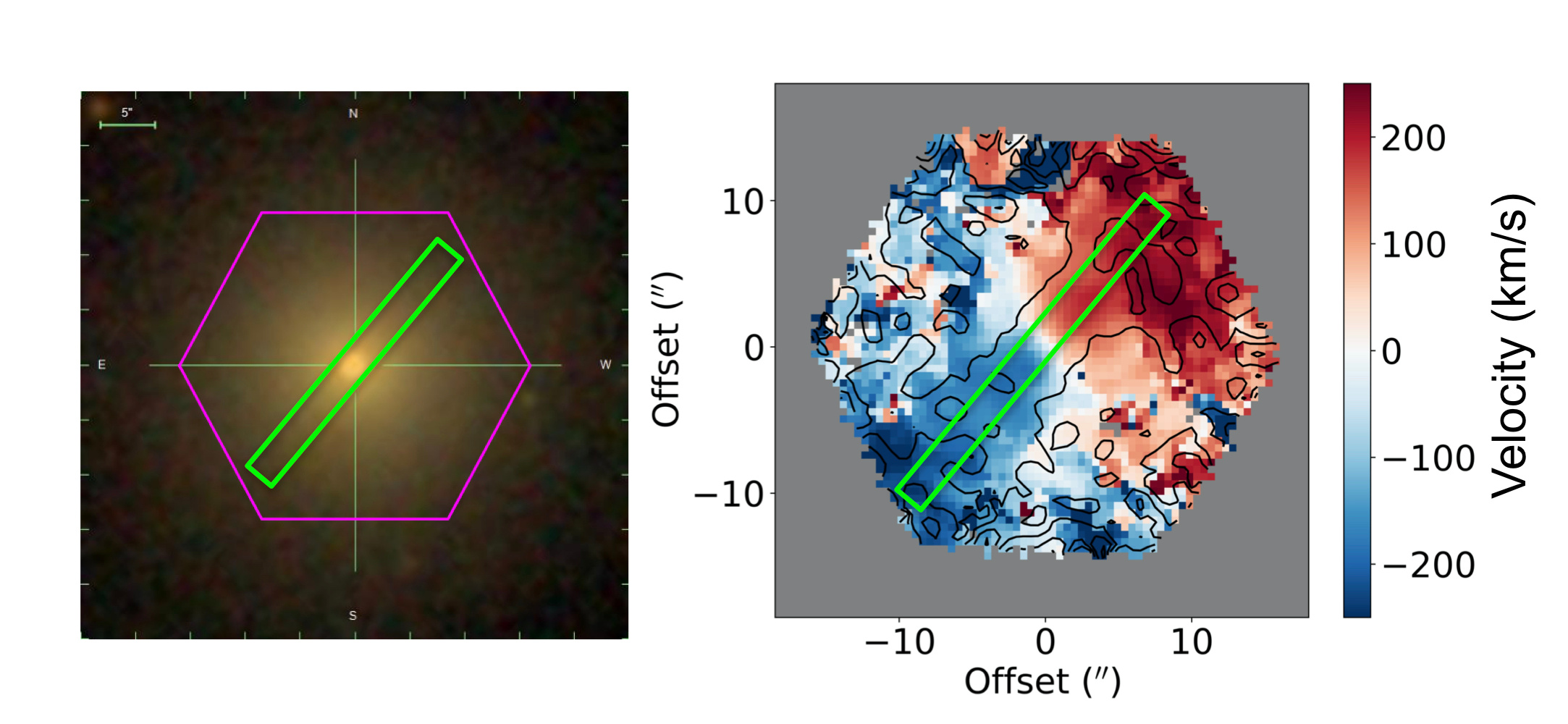}
\caption{  The ESI slit positions for the second target (MaNGA-ID: 1-145922). The left and the right panel show the slit position on the optical image and the MaNGA gas velocity map respectively. The black contours in the velocity map shows the H$\alpha$ EW. The slit is placed at an angle of 320$^{\circ}$ from North to East, tracing the bi-conical emission feature.
\label{fig:slit2}}
\end{figure}

Fig.~\ref{fig:slit1} and \ref{fig:slit2} show the slit positions for the two galaxies respectively, overlaid on their optical image (left panel) and on their MaNGA ionized gas velocity fields with H$\alpha$ EW contours on top (right panel). In the first target, slit 1 was placed at an angle of 40$^{\circ}$ from North to East tracing the bi-symmetric emission feature. Slit 2 was placed at an angle of 110$^{\circ}$ from North to East to sample the central parts of the galaxy and intersect with slit 1 at the center. Slit 3 ran parallel to slit 1 at an offset of $\sim\rm4.9''$ in the South-East direction. In the second galaxy, the slit was placed at an angle of 320$^{\circ}$ from North to East along its biconical pattern.

We used routines from the \textit{ESIRedux} package, developed by Jason X. Prochaska to aid in reducing the ESI data. This package was primarily built in IDL and is publicly available \footnote{ \textit{https://www2.keck.hawaii.edu/inst/esi/ESIRedux/}}. Our primary goal is to extract spatially resolved spectra along the multiple slits of our targets, whose angular extent on the sky fills the slit ($\sim20''$). This requires some steps in addition to the \textit{ESIRedux} pipeline. The detailed description of the data reduction steps is given in the next section.

\subsection{ESI data Reduction} \label{reduction}


We begin by using \textit{ESIRedux} to perform bias subtraction, flat field correction and wavelength calibration. Several dome flats are taken at the beginning of each night with the same instrument setup as the target observations. In the first step, the routine identifies and combines the bias and flat frames separately to create a median bias and a median flat frame. The median bias is subtracted from the image. The bias subtracted image is then normalized by the median flat. 
The resulting image is then run through the wavelength calibration routine in the pipeline. 

We utilize python routines to do the rest of the reduction steps. Since all the targets fill the 20$''$ ESI slit, object detection and sky spectra extraction can't be performed with \textit{ESIRedux} routines. Separate sky observations were taken near the target during observations roughly $\sim50\%$ of the time, alternatively switching between the sky and the targets. Sky subtraction is done by directly subtracting the sky image from the science image taken closest in time. We also account for small telescope pointing offsets that may arise between different exposures by cross-correlating and shifting the spectra from different exposure frames with respect to each other in order to be aligned perfectly. A 3$\sigma$ clipping routine is then applied to the sky subtracted science images to remove cosmic rays. Finally the individual exposures are added together to form a combined 2D image. Using the 2D wavelength solution obtained from wavelength calibration and the co-added 2D science image, we extract 1D spectra from each spatial pixel along the slit for each individual echellete order. Finally we stitch all ten echelette orders together to form a continuous spectrum over the entire wavelength range. To improve the signal-to-noise (SN) and ensure a minimum SN $>$ 1 over the entire extent of the slit, the spectra are binned spatially with a binsize of 1$''$.  

The final step of the data reduction is flux calibration. Standard star (G191b2b) observations taken at the beginning of each night have been used to calibrate the spectra to first order. Flux Calibration corrects initially for two effects: the blaze effect caused by the instrument response of the echelle grating and the conversion from photon counts to flux density in physical units. Both effects are corrected using standard star observations by comparing the absolute flux of the standard star to the observed counts sampled at the same wavelength. However, the science spectra, which have been flux-calibrated in this manner, show a systematic difference in flux values with that of the MaNGA spectra by non-negligible amount. The possible reason behind this mismatch maybe inaccurate blaze function removal, non-photometric conditions or variations along the slit. Since we already possess MaNGA spectra for our targets, with flux calibration accuracy to a few $\%$ \citep{yan16b}, we use the MaNGA spectra as a reference to correct for residual inaccuracies in the flux calibrated Keck spectra. 

To perform this second step flux calibration, we obtain the sky positions of each of the 1$''$ Keck ESI spatial bins (hereafter, spaxels) on the target galaxy. We then find the MaNGA spaxels which overlap in sky position, yielding 1-1 spatial mapping between the MaNGA spaxels (of size $\sim \rm 0.5 ''$) and ESI binned spaxels (1$''$). Next, for each ESI spaxel, we take the corresponding MaNGA spectrum (from the spatial mapping) and smooth it using a Gaussian filter. The smoothing uses a fairly large spectral window size while disregarding emission lines and other sharp spectral features. The smoothed MaNGA spectrum is then fit with a Chebyshev polynomial, the fit being given by the calibration vector, P$_{MaNGA}$. On the other hand, the corresponding ESI spectra is also smoothed by a gaussian filter, but using a window size almost 3 times as large, owing to the higher spectral resolution. The smoothed ESI spectrum is fit similarly with a Chebyshev polynomial which gives P$_{ESI}$. The final corrected flux calibrated ESI spectrum (ESI$_{final}$) is then obtained by scaling the original ESI spectrum (ESI$_{original}$) by the formula:
\begin{equation*}
    ESI_{final} = ESI_{original}\frac{P_{MaNGA}}{P_{ESI}}
\end{equation*}

\begin{figure}[h!!!] 
\centering
\graphicspath{{./plots/}}
\includegraphics[width = 0.5\textwidth]{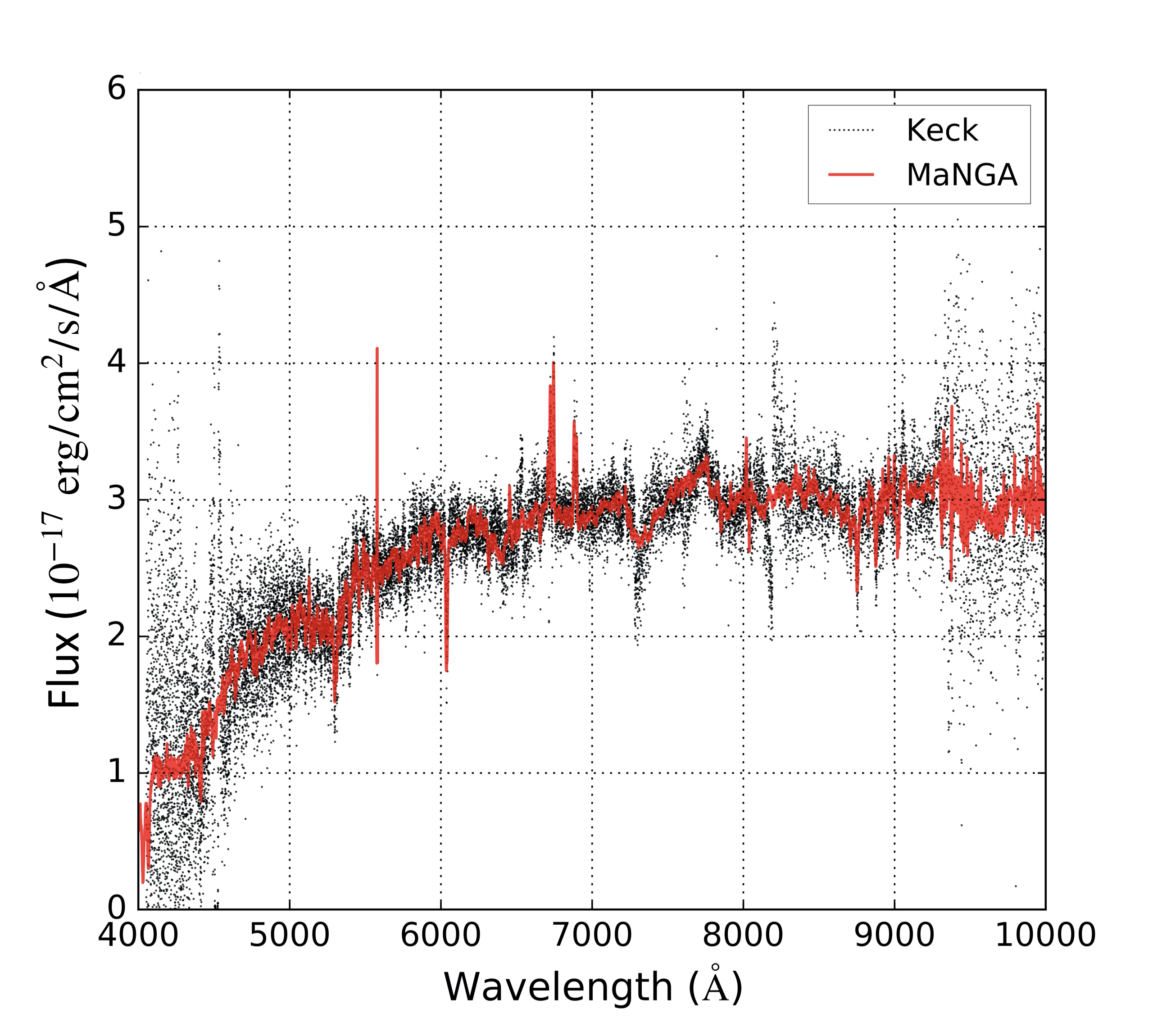}
\caption{  Example of a reduced Keck ESI spectrum (in black) in an 1$''$ spatial bin along the ESI slit. MaNGA spectra from the spaxel corresponding to the same physical position on the sky is overplotted in red. 
\label{fig:keck}}
\end{figure}

The above process is repeated for all spectra along each ESI slit. An example of a reduced ESI spectrum is shown in Fig.~\ref{fig:keck}. 
\\

\section{Method \& Analysis} \label{sec:method}

\subsection{Sample Selection} \label{sub:sample}

Our parent sample of red geysers consists of 84 galaxies which accounts for $\approx8\%$ of the quiescent MaNGA population \citep{roy18}. The red geyser sample is visually selected based on the following characteristic features:
\begin{itemize}[noitemsep]
\item Very low star formation rate with typical value of log SFR [$\rm M_\odot/yr] \sim -3$, measured by optical-IR SED fitting \citep{chang15, salim16}, and rest frame color $NUV-r>5$.
\item Bi-symmetric emission feature in spatially resolved H$\alpha$-EW 2-dimensional map.
\item Rough alignment of the bi-symmetric feature with the ionized gas kinematic axis, but misalignment with stellar kinematic axis.
\item High spatially resolved gas velocity values, typically reaching a maximum of $\rm~300~km~s^{-1}$, which are greater than the stellar velocity values by a few factors.
\end{itemize}

Further details about the full sample are given in \cite{roy18}.

\begin{figure*}[h!!!] 
\centering
\graphicspath{{./plots/}}
\includegraphics[scale = 1.6]{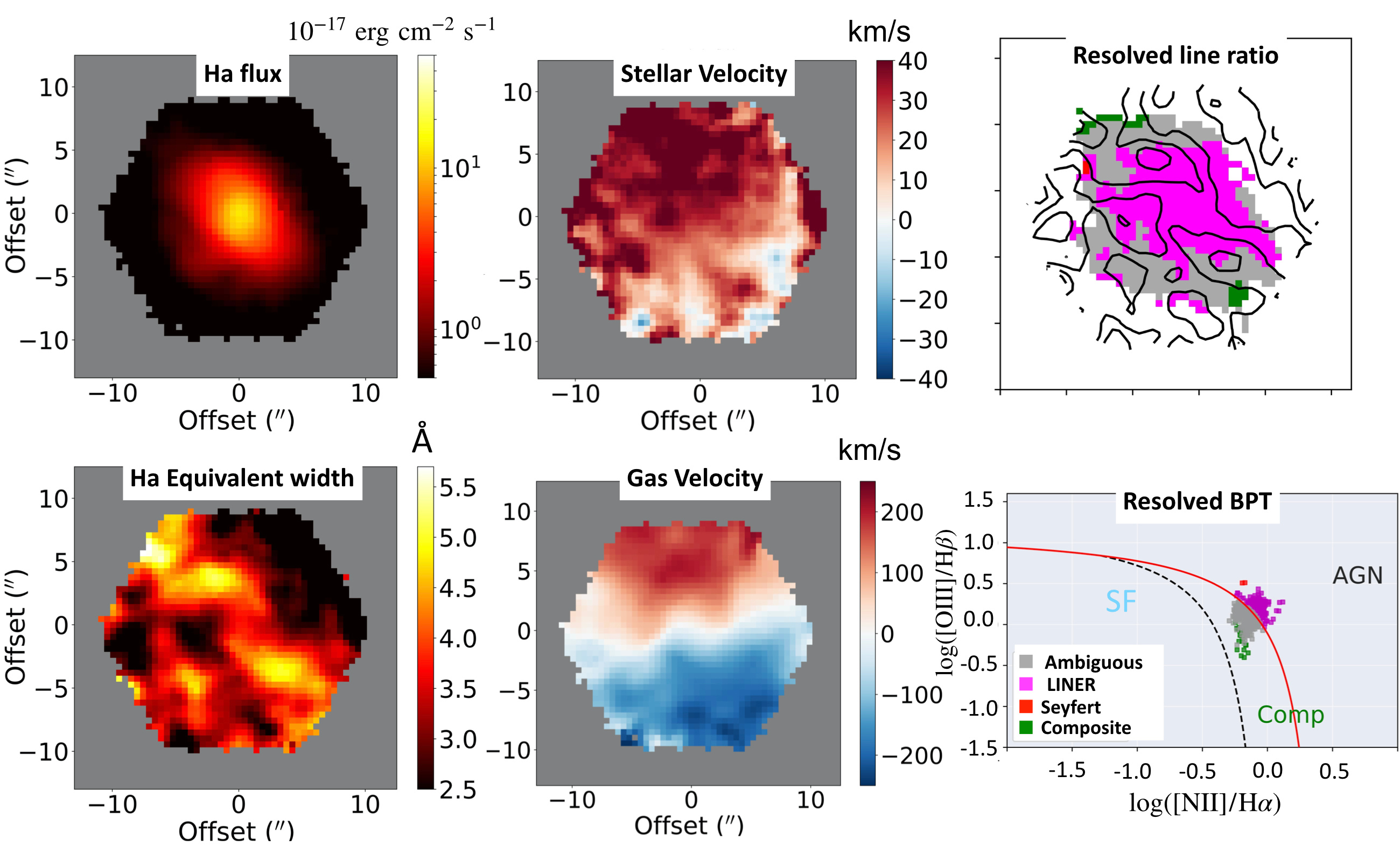}
\caption{ The spatially resolved emission line and kinematic properties of the first target red geyser galaxy from SDSS IV-MaNGA (MaNGA-ID: 1-217022). The on-sky diameter of this particular IFU fiber bundle is $\rm 22''$ which corresponds to a physical size of 11 Kpc. The upper left panel shows the spatial distribution of H$\rm\alpha$ flux. In the other panels, as labelled, we show the H$\alpha$-EW map (lower left), the 2D stellar velocity field (upper middle) and the ionized gas velocity traced by H$\alpha$ (lower middle). The lower right panel show the spatially resolved BPT diagram showing spaxels with signal to noise $>$ 3. The upper right panel shows the spatial position of those spaxels, colored by their classification based on the BPT diagram. Characteristic red geyser features such as the bi-symmetric emission feature in H$\alpha$ and its alignment with the gas kinematics axis are particularly apparent. 
\label{fig:prototype}}
\end{figure*}

\begin{figure*}[h!!!] 
\centering
\graphicspath{{./plots/}}
\includegraphics[scale = 1.6]{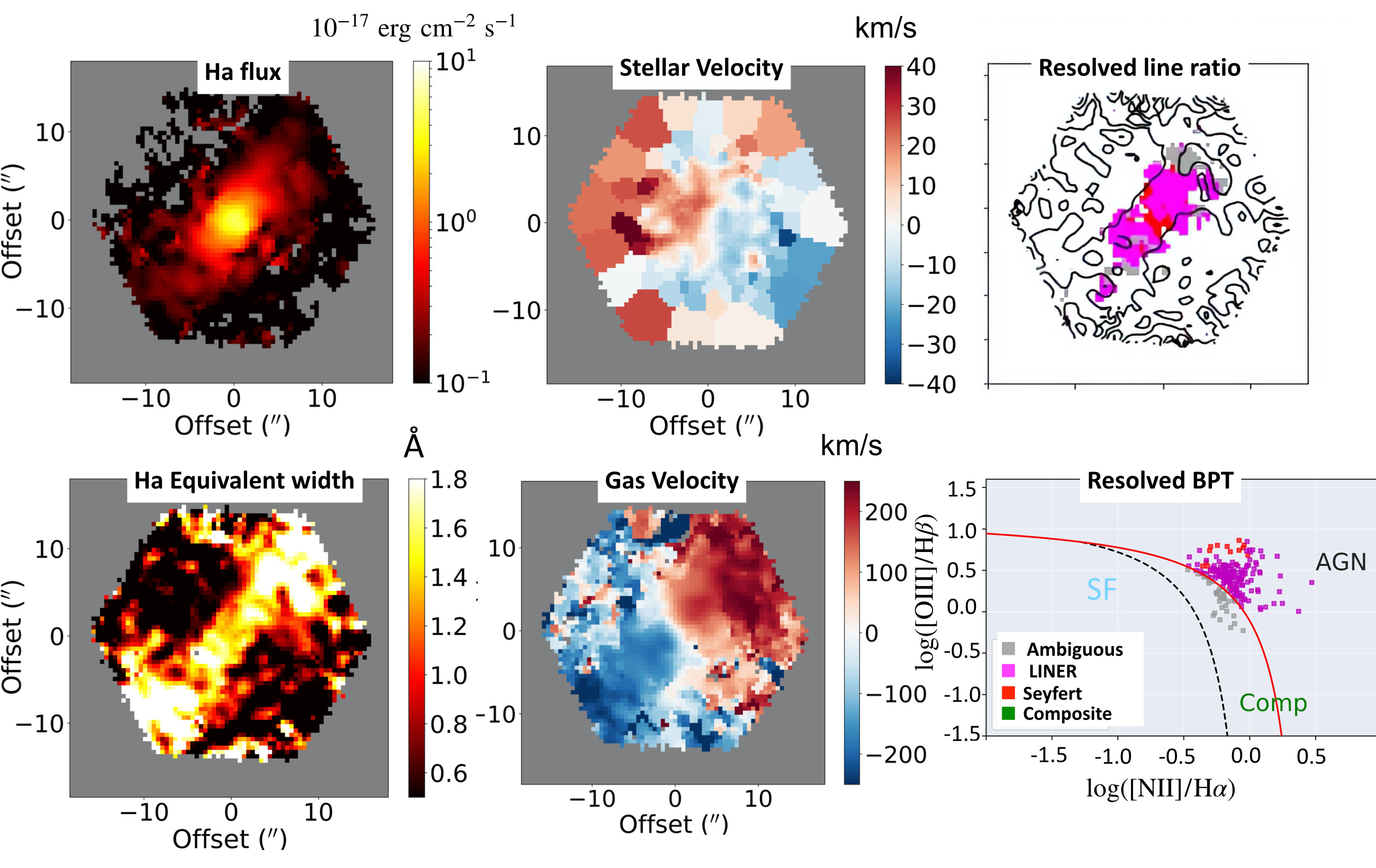}
\caption{ The spatially resolved emission line and kinematic properties of the second target red geyser galaxy from SDSS IV-MaNGA (MaNGA-ID: 1-145922). The on-sky diameter of this particular IFU fiber bundle is $\rm 32''$ which corresponds to a physical size of 18 Kpc. The different panels are the same as in Fig~\ref{fig:prototype}. 
\label{fig:prototype2}}
\end{figure*}

The two red geyser candidates selected for the Keck ESI followup represent a range of different values of H$\alpha$ flux, EW, ionized gas velocity, misalignment angle and radio flux while having sky positions amenable to the allocated Keck observing time. Hence they are good representations of the entire parent sample. The two chosen targets are shown in Fig.~\ref{fig:prototype} and \ref{fig:prototype2}.  
The first galaxy (ID: 1-217022) is the prototypical red geyser \citep{cheung16} which shows all the characteristics mentioned above and was among the first ones discovered in MaNGA. Its relevant characteristics are highlighted in Fig.~\ref{fig:prototype}. The top leftmost panel in the figure shows the H$\alpha$ flux distribution, which is extended in nature and shows a high value compared to other passive quenched galaxies, surpassing $\sim \rm 5~\times 10^{-16}~erg~cm^{-2}~s^{-1}~$. The galaxy has a clear bi-symmetric pattern in the H$\alpha$ EW map with an average EW $\sim$4.5~\AA~(bottom left panel). The top and bottom middle panels show the stellar and gas velocity maps respectively. \cite{cheung16} performed detailed dynamical modeling of the gas and stellar kinematics in this galaxy to conclude that a centrally driven outflow is the likely explanation for the high ($\rm \sim~300~km~s^{-1}$) gas velocity values. The spatially resolved BPT diagrams (rightmost panels) show LINER-like line ratios through out the galaxy. This galaxy has a central radio source detected in FIRST and Jansky Very Large Array (JVLA) 1.4 GHz continuum emission \citep{cheung16}, which indicates a low-luminosity radio AGN with low Eddington ratio ($\lambda \sim \rm 10^{-4}$). The presence of ionized gas is evident in the strong detection of Balmer lines and forbidden emission lines like H$\alpha$, [NII] and [OIII]. 

The second red geyser target (ID: 1-145922) also shows a clear bi-symmetric EW pattern (Fig.~\ref{fig:prototype2}), but with a much lower H$\alpha$ EW value $\sim \rm~ 1 - 2.5 $~\AA~(bottom left panel). LINER-like line ratios are observed through most of the galaxy while AGN/Seyfert line ratios appear in the center (right panels), with log([$\rm NII]/H\alpha$) > $-$0.4. Unlike the first target, this galaxy doesn't show systematic rotation in the stellar kinematics (upper middle panel) and hence the exact misalignment angle between the gas and stellar velocity field is unclear. However, the estimated average second velocity moment ($V_{\rm rms}\equiv \sqrt{V^2 + \sigma^2}$) of the ionized gas and stars are similar to the first target. 
This galaxy is not radio detected in the FIRST survey and has not been followed up with JVLA.  
Both galaxies have spheroidal morphology as observed from SDSS ground based imaging. 

\begin{figure}[h!!!] 
\centering
\graphicspath{{./plots/}}
\includegraphics[width = 0.5\textwidth]{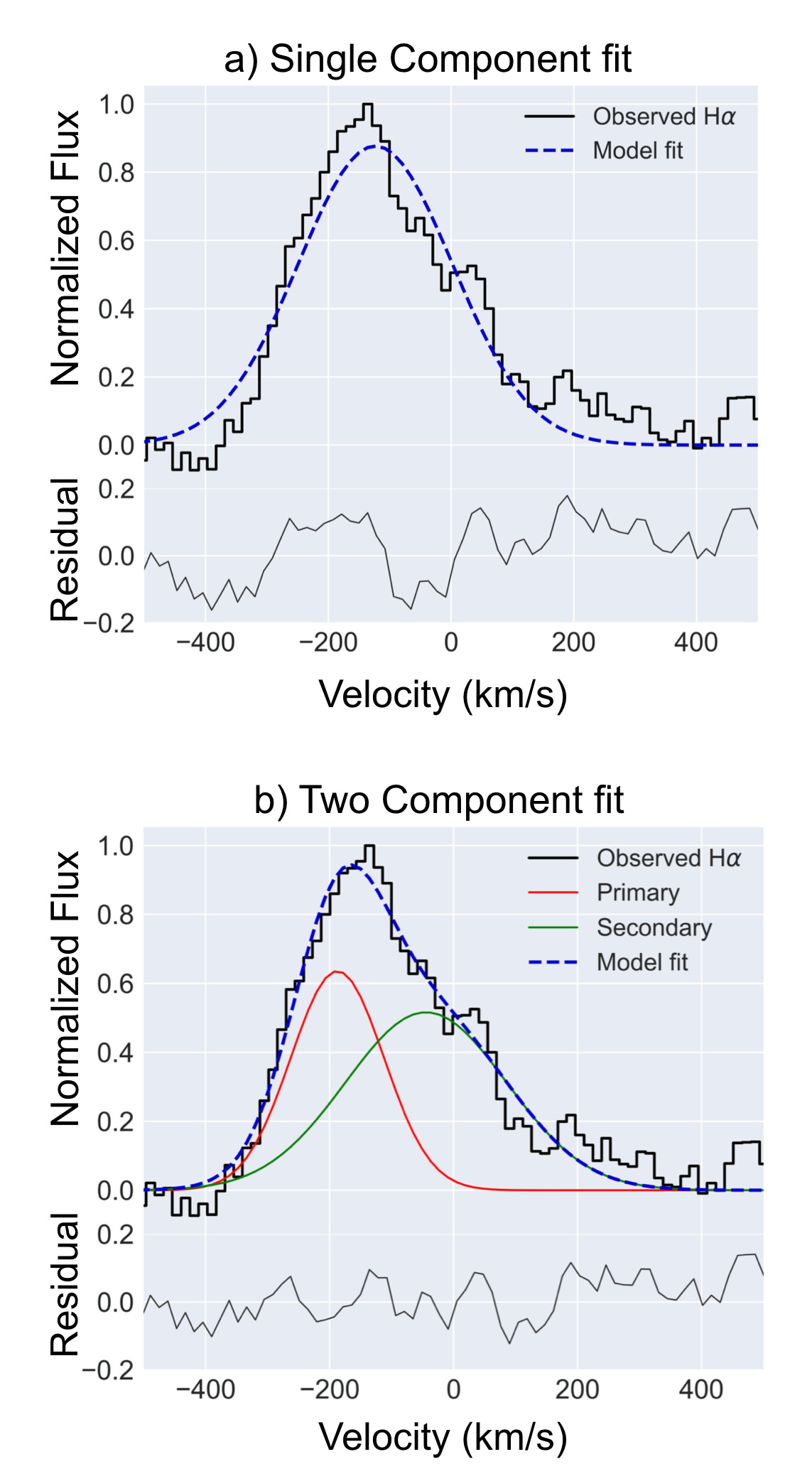}
\caption{ [\textit {Top panel}] An example of a single gaussian fit to the continuum subtracted H$\alpha$ emission line from ESI at a particular 1$''$ spaxel. Single component model (shown in blue) leaves huge residual of about $\rm\pm 15\%$ as shown in the residual panel. [\textit {Bottom panel}] A two-component model fit to the same spectrum as above is shown. The model consist of a double gaussian, with the ``Primary'' component shown in red measuring the bulk velocity of the gas, and the ``Secondary'' component shown in green which shows the wings and departures from single component model. This model greatly reduces the residual to only about $\rm\pm 5\%$ and is favored. 
\label{fig:fit}}
\end{figure}

\begin{figure}[h!!!] 
\centering
\graphicspath{{./plots/}}
\includegraphics[width = 0.5\textwidth]{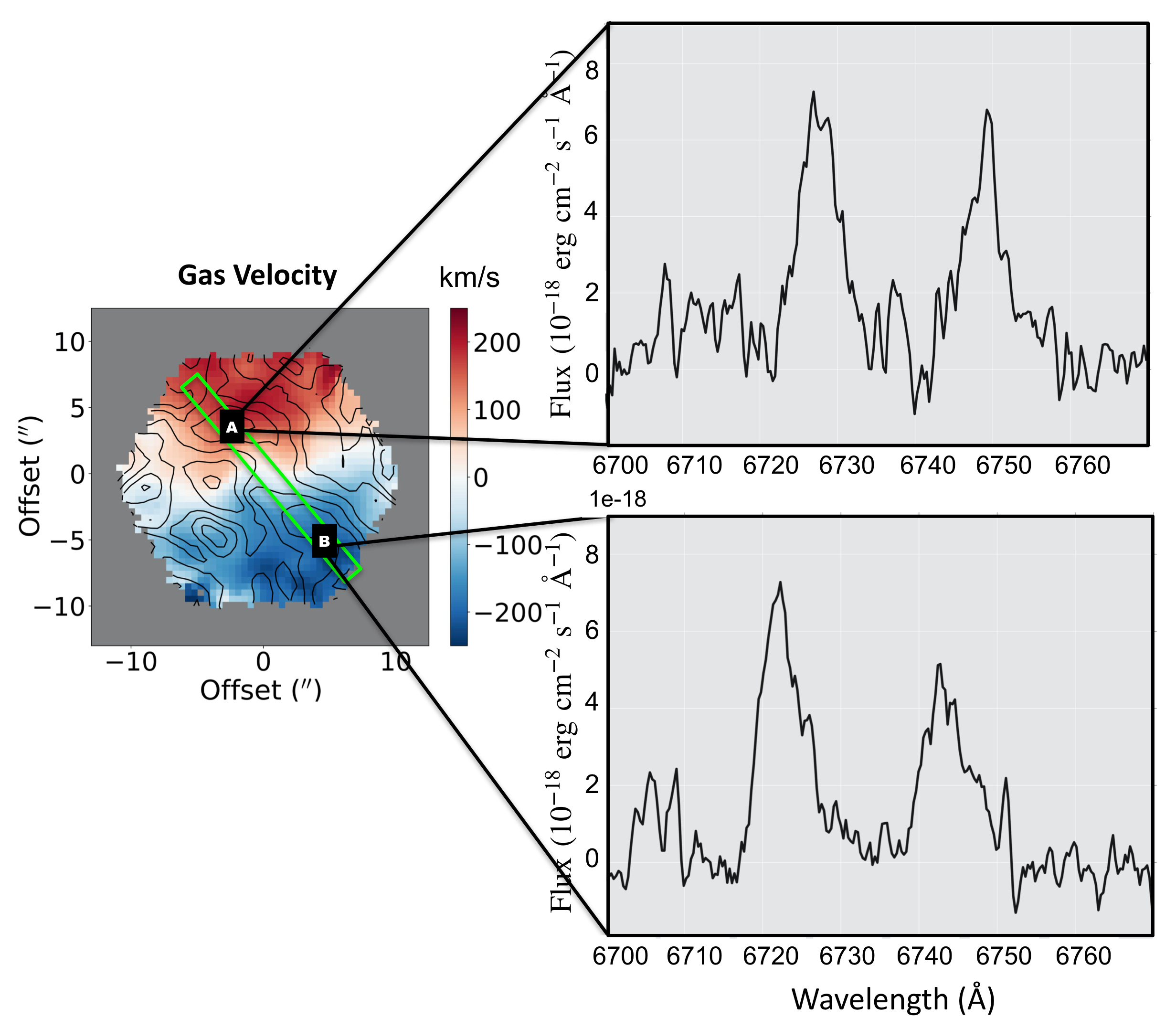}
\caption{  [\textit {Left panel}] The 2D gas velocity field of the first target galaxy obtained from MaNGA data with H$\alpha$-EW contours overplotted in black. [\textit {Right panel}] H$\alpha$ and [NII]$\lambda6584$ emission line profiles obtained from Keck ESI at two separate locations in the galaxy indicated A and B in the left panel. The emission line profiles reveal complex velocity structure likely resulting from multiple components. Extended wings in the emission lines switch from the blue side at position A to the red side at position B.
\label{fig:eg}}
\end{figure}

\subsection{Line profile fitting} \label{fitting}

An important aspect of accurately measuring emission lines is properly accounting for the stellar continuum. This is particularly important for the Balmer lines since underlying stellar absorption can lead to incorrect emission line flux and equivalent width estimation. 
We apply MaNGA's data analysis pipeline (DAP), version 2.3.0 \citep{westfall19} on the fully reduced Keck data for accurate modeling of the stellar continuum throughout the galaxy. First, we run DAP modules on each spaxel in the MaNGA datacube of the target galaxy. The DAP masks $\pm\rm750~km~s^{-1}$ surrounding each of the expected emission line regions at the galaxy systemic redshift and fits the stellar continuum using the Penalized Pixel fitting algorithm {\tt pPXF} \citep{cappellari04, cappellari17}. It uses a combination of 49 templates based on the MILES stellar library, known as MILES-HC library \citep{MILES}, which provides statistically equivalent fits to those that use the full library of 985 spectra in the MILES stellar library \citep{westfall19}. Once the DAP fits the continuum and obtains the optimal stellar template for a particular MaNGA spaxel, we use that same template to fit the continuum of the Keck spectra for the corresponding Keck spaxel, again using {\tt pPXF}. The best fit stellar continuum is then subtracted from the observed spectrum before we analyze the emission lines.

We treat H$\alpha$ and [NII]$\lambda\rm~6584$~\AA~emission lines independently, fitting for their flux, velocity and dispersion. At first we model both the $\Ha$ line and [NII]$\lambda\rm~6584$~\AA~line as  single Gaussian profiles. After binning the ESI slit data into spatial bins of size 1$''$, we constrain the velocity within each bin to be within $\rm\pm~350~km~s^{-1}$ of the systemic velocity. This is done to prevent the module from fitting spurious peaks. We also require the dispersion to be less than 500$\rm ~km~s^{-1}$. In order to sample the entire parameter space and to obtain an unbiased estimate of the uncertainties in the fitted values, we wrap our fitting procedure in a Markov Chain Monte Carlo (MCMC) framework with Dynamical Nested Sampling algorithm \citep{higson19} using the Python package \textit{dynesty} \citep{speagle19}. The nested sampling method \citep{skilling06, skilling04} is used to estimate both the Bayesian evidence and posterior distribution in an iterative fashion until the convergence criteria is met. However, the single Gaussian model
results in large residuals for most of the spatial bins (or spaxels). Fig.~\ref{fig:fit} panel a (top) shows the single Gaussian fit of the spectrum around H$\alpha$ emission line for a particular ESI spaxel in the first target galaxy (MaNGA-ID: 1-217022). It is clear from the figure that the emission line show large departure ($\pm \ 15 \%$) from the model. 
We therefore proceed to fit a double Gaussian model to the emission lines, keeping the velocity of the first Gaussian component within $\rm\pm~50~km~s^{-1}$ of the velocity estimate obtained from single component fitting. This ``primary component'' is constrained to have a greater flux than the ``secondary component''. The dispersion in the primary component is constrained to be $\rm < 200~km~s^{-1}$ while no restriction is imposed on the secondary component. This is done to make sure that the primary component, or the ``narrow'' component, represents the bulk velocity of the gas while the secondary component is sensitive to broad wings and other departures from the primary.  Fig.~\ref{fig:fit}, panel b (bottom) shows the double Gaussian fit of the same spectrum, with primary component shown in red and secondary component in green. The resulting fit (blue dashed line) shows that the flux residuals are less than $\rm5\%$ of the median flux.

To determine which of the two models is preferred for each spaxel, we use a Bayesian Information Criterion (BIC) which uses the likelihood to determine the optimum model but penalizes for additional free parameters. It is defined as : 

\begin{equation}
    \rm BIC = -2 \mathcal{L} + k~log (N)
\end{equation}

where $\mathcal{L}$ is the log-likelihood, k is the number of free parameters and N is the number of data points that get fit. The model with the lower BIC value is the preferred model and the final fit is selected accordingly. For example in Fig.~\ref{fig:fit}, the double Gaussian model is preferred over the other.



    


\section{Results} \label{results}

With emission line velocity profiles extracted from the higher resolution Keck ESI data and fit with single or double Gaussian model depending on the BIC criterion, we are ready to investigate systematic patterns in the profile shape emerging in different parts of the galaxy. An example of H$\alpha$ and [NII] emission line profiles in two opposite ends of the galaxy is shown in Fig.~\ref{fig:eg}. This detailed line profile information may inform or refute our earlier wind model interpretation of red geyser gas kinematics that was based on MaNGA data alone.  We begin by presenting all the Ha and [NII] velocity profiles derived from the ESI data at different slit positions for both of the target galaxies.

\subsection{Target 1-217022} \label{target1}

For the first target, slit 1 is placed at a position angle of 40$^{\circ}$ from North to East along the bi-symmetric emission feature. The observed spectra are shown in Fig.~\ref{fig:spectra4}. The top panel shows the MaNGA 2D ionized gas velocity field of the target, with H$\alpha$ EW contours overplotted in black. The slit position is overlaid on top with white filled circles indicating the different spaxels of 1$''$ where reduced spectra have been extracted. In the bottom panel, the different rows indicate those reduced spectra with H$\alpha$ (emission line on the left side) and [NII]$\lambda6584$ (on the right) lines shown for each 1$''$ spaxel of physical size of 0.51 kpc along the slit. The spectra are color coded by the velocity estimate obtained by fitting single Gaussian model to the H$\alpha$ emission line. The spectra are then shifted along the wavelength axis so that the emission lines from the different spaxels are roughly aligned. This is done to enable easy visual comparison of the shapes and asymmetry of the emission lines.

The gas velocity amplitude rises steeply from the center outwards both on the red and blue-shifted side. The details in the emission line profiles are shown in Fig.~\ref{fig:multigauss4} via Gaussian model fits of H$\alpha$ and [NII] lines at different spaxels (spaxel number shown in the upper right or left corner in each subplot). The primary velocity component reflects the bulk velocity of the ionized gas. If the double Gaussian model is preferred (according to the BIC criterion as mentioned in \S\ref{fitting}), the line profile is asymmetric and wings or any departures from the bulk velocity is then reflected in the secondary velocity component. The best fit parameters for all spaxels are given in Appendix \ref{appen:fitparam}.

We find that the primary velocity component (shown in red for the 2-component model and in blue for the one component model, Fig.~\ref{fig:multigauss4}) goes from blue-shift $\rm\sim-250~km~s^{-1}$ from one end of the slit to a redshift of $\rm\sim250~km~s^{-1}$ in the other closely matching the values observed from the MaNGA velocity map (Fig.~\ref{fig:spectra4}, top panel). The secondary fainter component (in green) varies widely in full-width half maxima (FWHM; 2.355$\rm\cdot\sigma$) ranging from $\rm\sim150-600~km~s^{-1}$, with broad wings near the center and the red-shifted side of the galaxy. On average, emission lines with a blue-shifted primary velocity component show a secondary component shifted redward. The secondary component switches to the bluer side when the primary velocity component is red-shifted. In other words we see a ``red'' wing on the blue side of the galaxy and a ``blue'' wing on the red side. Infact, for this particular slit position, the secondary component in certain spaxels have shifted so much relative to the primary that its velocity relative to the systemic velocity have opposite signs compared to the primary component. That means, on the red-shifted side, the blue wing is actually blue-shifted relative to the systemic velocity of the galaxy and vice versa on the other side of the galaxy. The line profiles show mostly symmetric profiles near the galaxy center where a single Gaussian model is preferred. The implication of these findings are discussed in \S\ref{sec:discussion}.  


For the second slit configuration (Fig.~\ref{fig:spectra3}), the slit lies almost perpendicular to the bi-symmetric feature, at an angle of 110$^{\circ}$ from North to East. From the H$\alpha$-EW 2D map, it is clear that the EW of the ionized emission is enhanced along the bi-symmetric feature, hence there is comparatively less signal perpendicular to it. The spectra extracted in different spaxels along the second slit is shown in the bottom panel of Fig.~\ref{fig:spectra3}. Near the edges of the slit, line-fitting becomes unsuccessful owing to low signal-to-noise.
The bulk velocity in slit 2 varies from $\rm\sim50~km~s^{-1}$ to $\rm\sim-50~km~s^{-1}$ (Fig.~\ref{fig:multigauss3}). The emission lines along this slit are substantially broad, with FWHM $> \rm 250~km~s^{-1}$ going up to $\sim \rm 600~km~s^{-1}$, similar to a few cases in slit 1. The average FWHM of [NII] lines is higher than that of H$\alpha$ and broad lines (with FWHM $>\rm300~km~s^{-1}$) are observed reaching a distance of \textit{$\sim1$} kpc from the center along the slit. The observed emission lines are generally symmetric across slit 2. Very slight asymmetry occasionally arises near the edges.



The third slit (Fig.~\ref{fig:spectra2}) is placed parallel to the first slit, i.e. parallel to the bi-symmetric feature, but slightly off-centered at a distance of $\sim\rm2.5$ kpc in the South-East direction. 
Owing to the decrease of ionized emission in the outskirts of the galaxy, the signal to noise of the spectra drops overall, compared to the first slit. Nevertheless, the line profile clearly shows a red wing on the blue side of the galaxy and a blue wing on the red side similar to slit 1 (Fig.~\ref{fig:multigauss2}). The degree of asymmetry is less, but the primary velocity component goes from blue-shift $\rm\sim-200~km~s^{-1}$ on one end of the slit to a redshift of $\rm\sim250~km~s^{-1}$ on the other. Just like slit 1, there is a significant velocity offset ($\sim \rm 100 - 150~km~s^{-1}$) between primary and secondary components. We note that in the MaNGA H$\alpha$-EW contour map, there is an additional H$\alpha$ enhanced region $\sim\rm 2.9$ kpc south east of the center. The overlapping ESI spectra in the 3rd row from the bottom (Fig.~\ref{fig:spectra2}, bottom panel) shows broader H$\alpha$ and [NII] lines possibly due to a distinct velocity component associated with this region.

\subsection{Target 1-145922} \label{target2}

The second target has only one slit observation (Fig.~\ref{fig:spectra}) with sufficient signal-to-noise spectra (a minimum SN $>$ 1 across the slit). The slit is oriented along the bi-symmetric feature, at a position angle of 320$^{\circ}$ from North to East. Despite the lower signal-to-noise, the fitting procedure was successful in extracting individual velocity components (Fig.~\ref{fig:multigauss}). The bulk velocity varies from $\rm\sim-250~km~s^{-1}$ to about $\rm\sim300~km~s^{-1}$. The average FWHM is lower compared to the first target with an average value between $\rm\sim150-200~km~s^{-1}$, but it reaches a maximum value of over $\rm\sim600~km~s^{-1}$ near the center. The blue-shifted part of the galaxy exhibits a secondary component with a velocity $\rm\sim150-200~km~s^{-1}$ redder than the primary velocity. This component provides an obvious ``red'' wing on the blue side. The profiles become symmetric towards the center before showing a blue wing in the red-shifted side.


\subsection{Measured velocity asymmetries } \label{result2}

In order to measure the observed asymmetry of the emission lines, we fit the emission lines with a Gauss-Hermite polynomial of the form:

\begin{equation}
   \rm  f(x) = Ae^{\frac{-g^2}{2}} [ 1+h_3(-\sqrt{3}g+\frac{2}{\sqrt{3}}g^3 ) + h_4 (\frac{\sqrt{6}}{4}-\sqrt{6}g^2+\frac{\sqrt{6}}{3}g^4)]
\end{equation}
\\

where $\rm g = \frac{x-x_c}{\sigma}$. Here A is the flux, $\rm x_c$ is the peak wavelength and $\sigma$ is the velocity dispersion. Departures from symmetry are quantified by the coefficient $\rm h_3$, which is a proxy for the skewness parameter ($k$). A positive $k$ parameter indicates the presence of a ``red'' wing. Similarly blue-winged components have a negative $k$ value. $ k = 0$ signifies a perfectly symmetric profile. 
To estimate the errors on the measured skewness, we perform MCMC simulations of the above model fit for every spaxel. We construct one hundred realizations of each spectrum by adding Gaussian noise with amplitude comparable to the noise measured in the original spectrum and quote the standard deviation of the $k$ parameter distribution, thus obtained, as the 1$-\sigma$ uncertainty on the measured skewness.

\begin{figure}[h!!!] 
\centering
\graphicspath{{./plots/}}
\includegraphics[width = 0.5\textwidth]{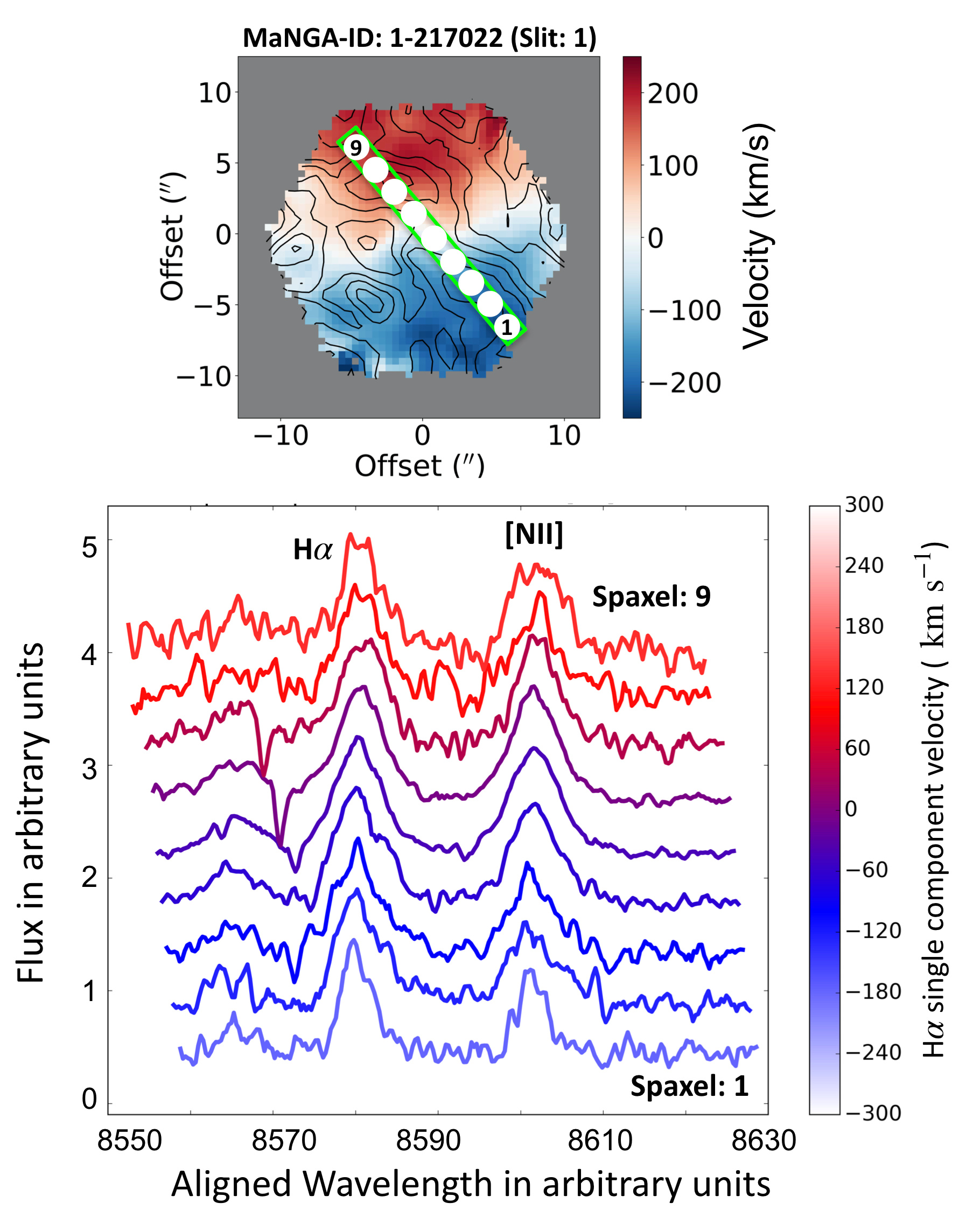}
\caption{  [\textit {Top panel}] The 2D ionized gas velocity field of the first target galaxy obtained from MaNGA data with the H$\alpha$ EW contours overplotted in black. The first slit position with different spatial bins ($\sim 1''$) is overlaid on top to map the physical location of the extracted spectra on the slit. [\textit {Bottom panel}] The different rows indicate the observed spectra from Keck ESI around the spectral window of H$\alpha$, [NII]$\rm \lambda 6584$ extracted from  different spatial bins (or spaxels) as marked in the top panel. The spectra are registered to the same wavelength and color coded by gas velocity.
\label{fig:spectra4}}
\end{figure}

\begin{figure}[h!!!] 
\centering
\graphicspath{{./plots/}}
\includegraphics[width = 0.5\textwidth]{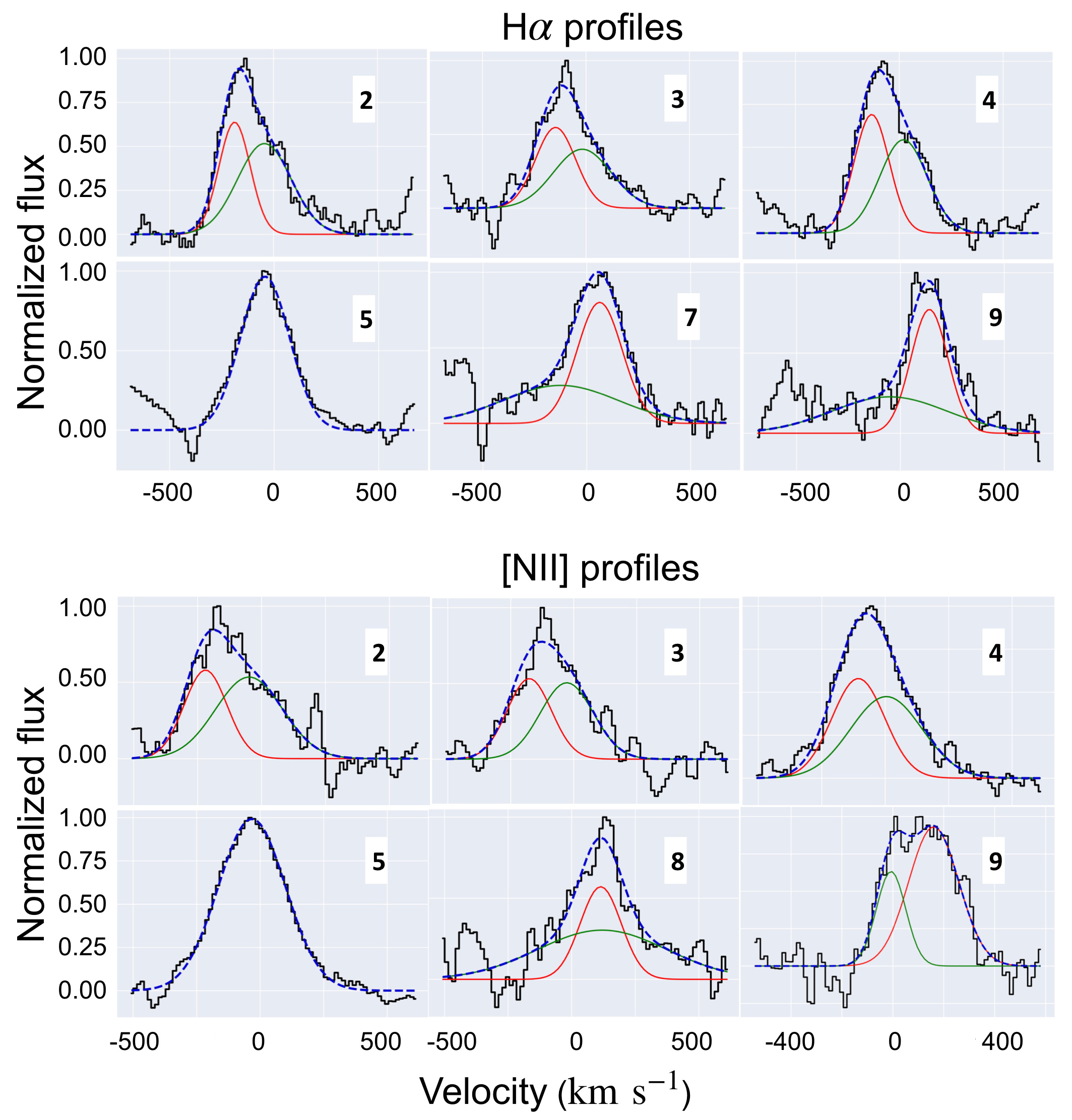}
\caption{[\textit {Top panel}] The H$\alpha$ emission lines from ESI for slit 1 position of the first target galaxy are shown for different spaxels along the slit in different panels. A single or a double Gaussian model, as favored by the Bayesian Information criterion (BIC), is used to model the emission lines. For a single component model the model is shown in blue. For two component models, the primary component is shown in red and secondary in green. The modelled line profile in blue shows the total model fit to the data (which is in black). [\textit {Bottom panel}] The [NII]$\rm \lambda 6584$ emission line profiles in black with their model fits over-plotted in blue, similar to the sub-panel above. 
\label{fig:multigauss4}}
\end{figure}

\begin{figure}[h!!!] 
\centering
\graphicspath{{./plots/}}
\includegraphics[width = 0.5\textwidth]{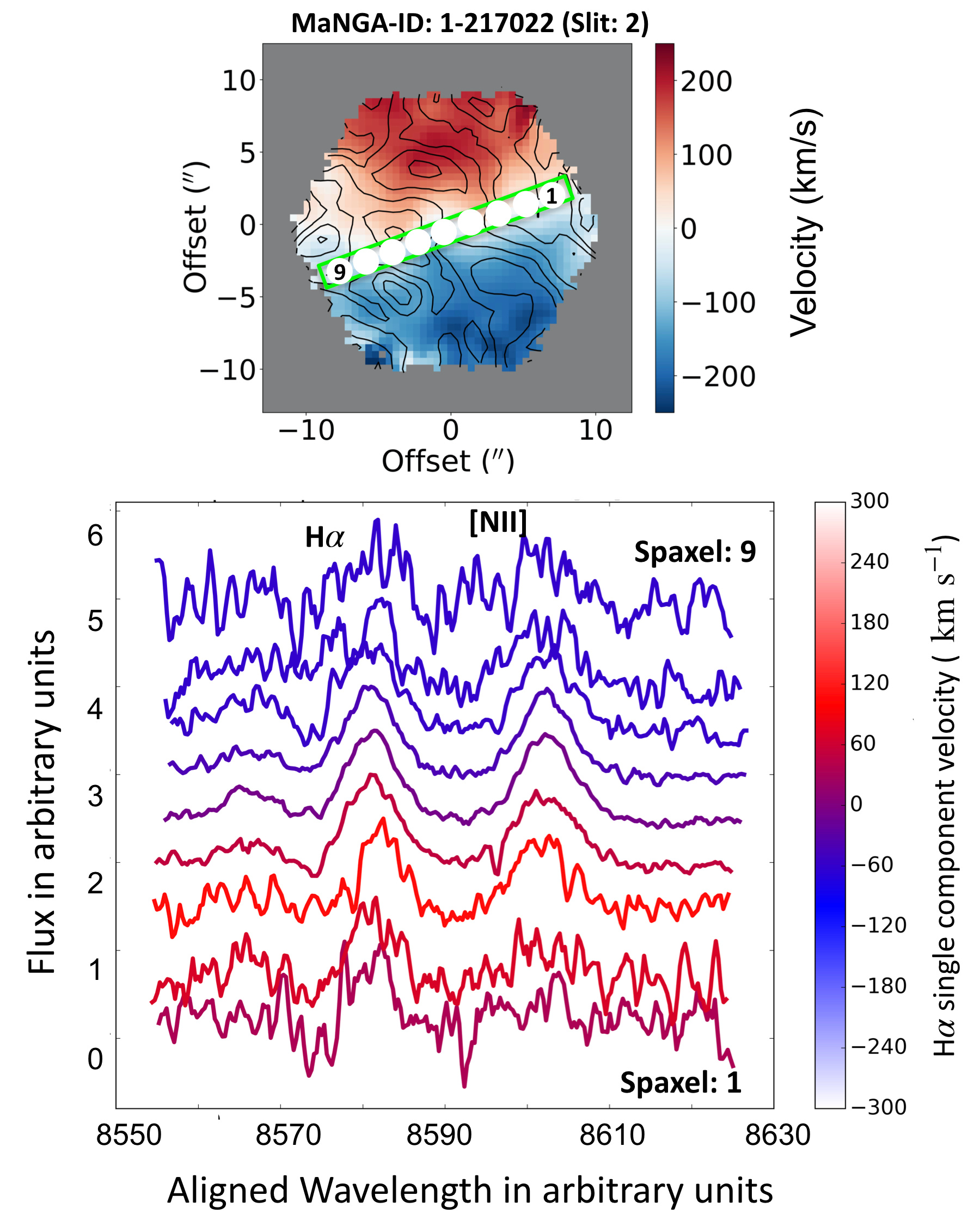}
\caption{ [\textit {Top panel}] The MaNGA ionized gas velocity field with H$\alpha$ EW contours overplotted in black, similar to Fig.~\ref{fig:spectra4}. The second slit position with different spaxels ($\sim 1''$) are overlaid on top. [\textit {Bottom panel}] The different rows indicate the observed spectra from Keck ESI around the spectral window of H$\alpha$, [NII]$\rm \lambda 6584$ extracted from  different spaxels as marked in the top panel. The spectra are registered to the same wavelength and color coded by gas velocity. 
\label{fig:spectra3}}
\end{figure}

\begin{figure}[h!!!] 
\centering
\graphicspath{{./plots/}}
\includegraphics[width = 0.5\textwidth]{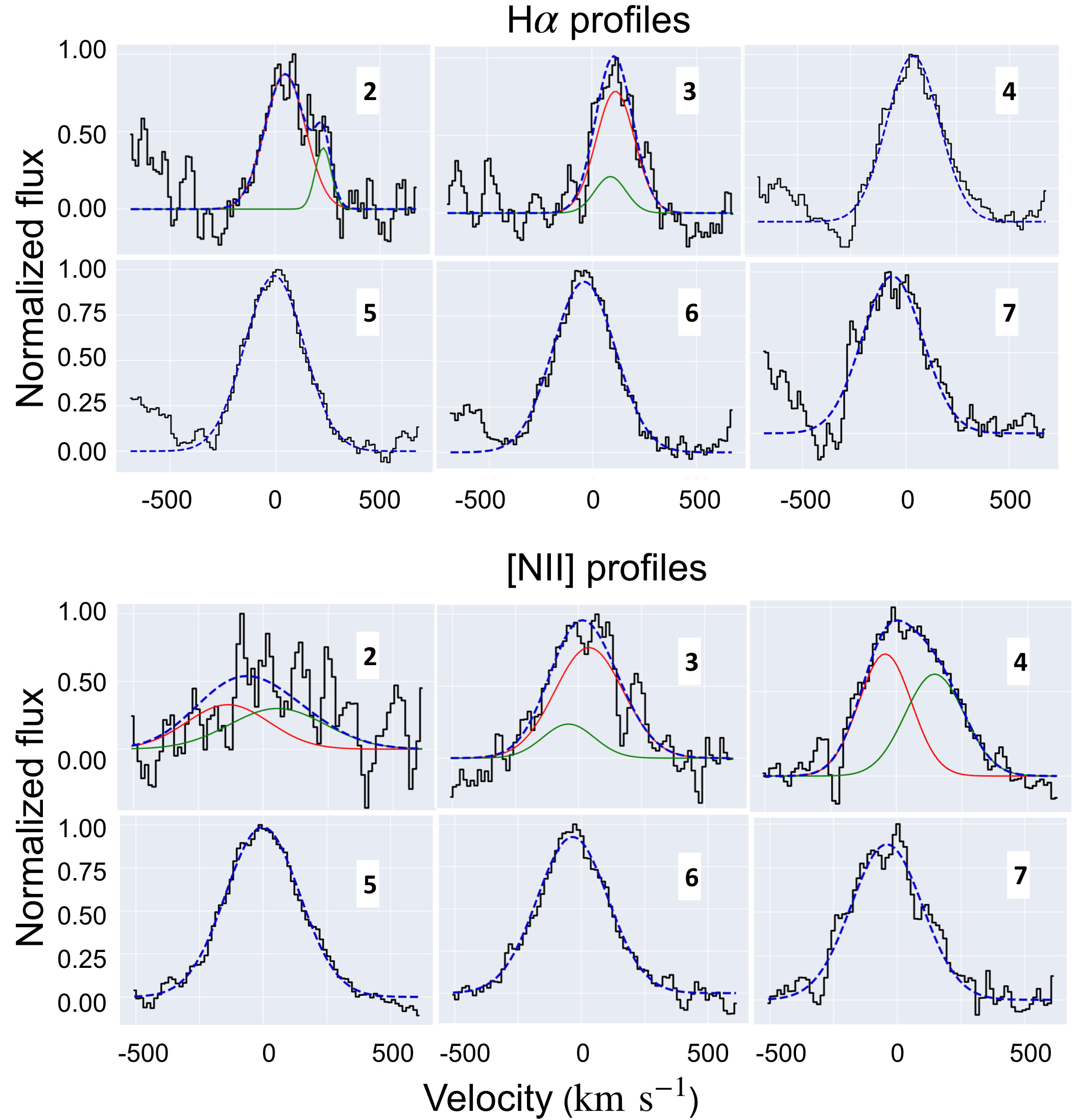}
\caption{[\textit {Top panel}] The H$\alpha$ emission lines from ESI for slit 2 position of the first target galaxy are shown for different spaxels along the slit in different panels. A one or two component Gaussian model is used to model the emission lines similar to Fig.~\ref{fig:multigauss4}. The modelled line profile in blue shows the total model fit to the data (which is in black). [\textit {Bottom panel}] Model fit for [NII]$\rm \lambda 6584$ emission line profiles, similar to the sub-panel above. 
\label{fig:multigauss3}}
\end{figure}

\begin{figure}[h!!!] 
\centering
\graphicspath{{./plots/}}
\includegraphics[width = 0.5\textwidth]{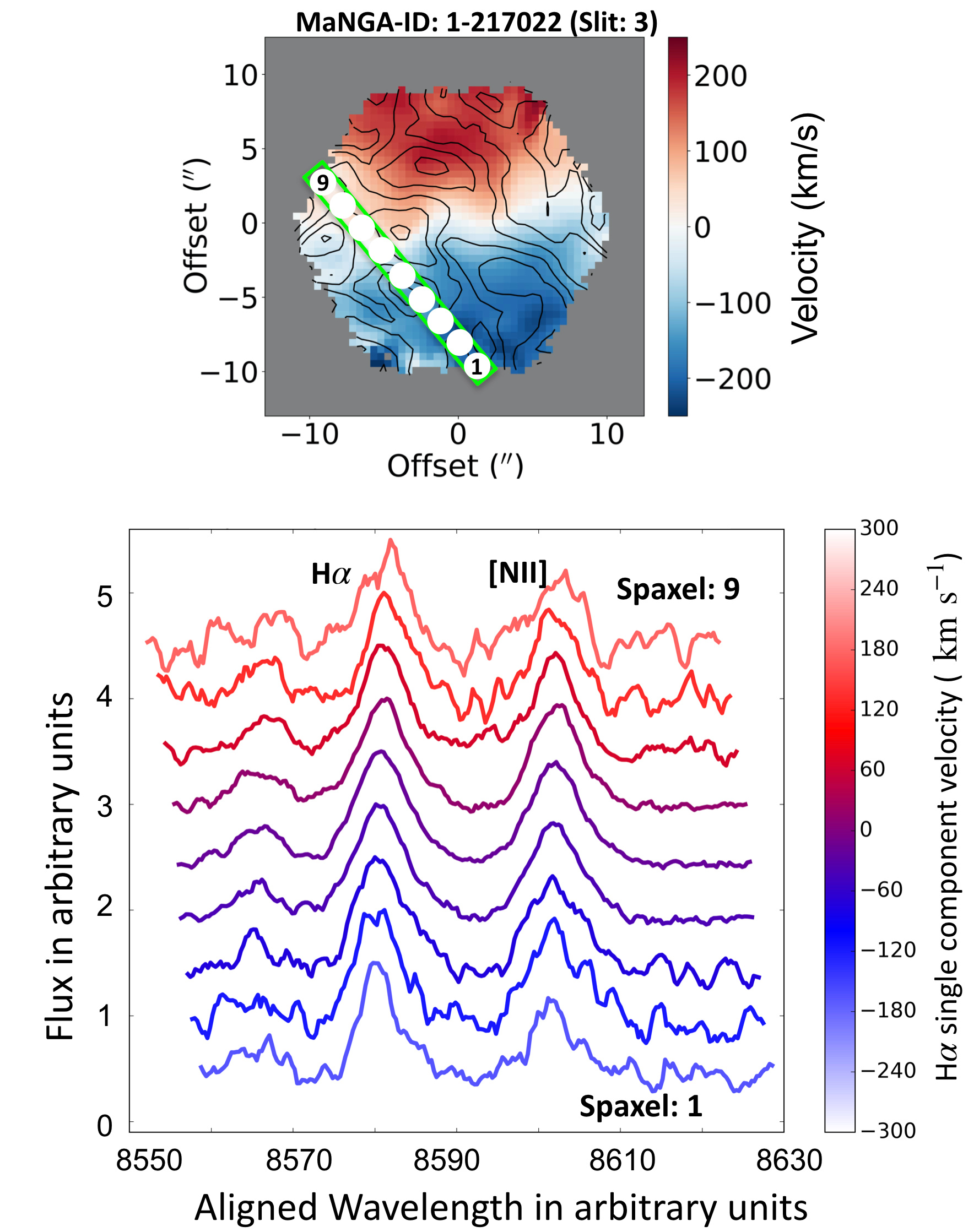}
\caption{[\textit {Top panel}] The MaNGA ionized gas velocity field with H$\alpha$ EW contours overplotted in black, similar to Fig.~\ref{fig:spectra4} $\&$ \ref{fig:spectra3} for the third slit position. This slit is parallel to but offset to the bi-symmetric feature, hence traces the edge of the putative wind cone. [\textit {Bottom panel}] The different rows indicate the observed H$\alpha$, [NII]$\rm \lambda 6584$ emission lines from Keck ESI extracted from the spaxels marked in the top panel. The spectra are registered to the same wavelength and color coded by gas velocity. 
\label{fig:spectra2}}
\end{figure}

\begin{figure}[h!!!] 
\centering
\graphicspath{{./plots/}}
\includegraphics[width = 0.5\textwidth]{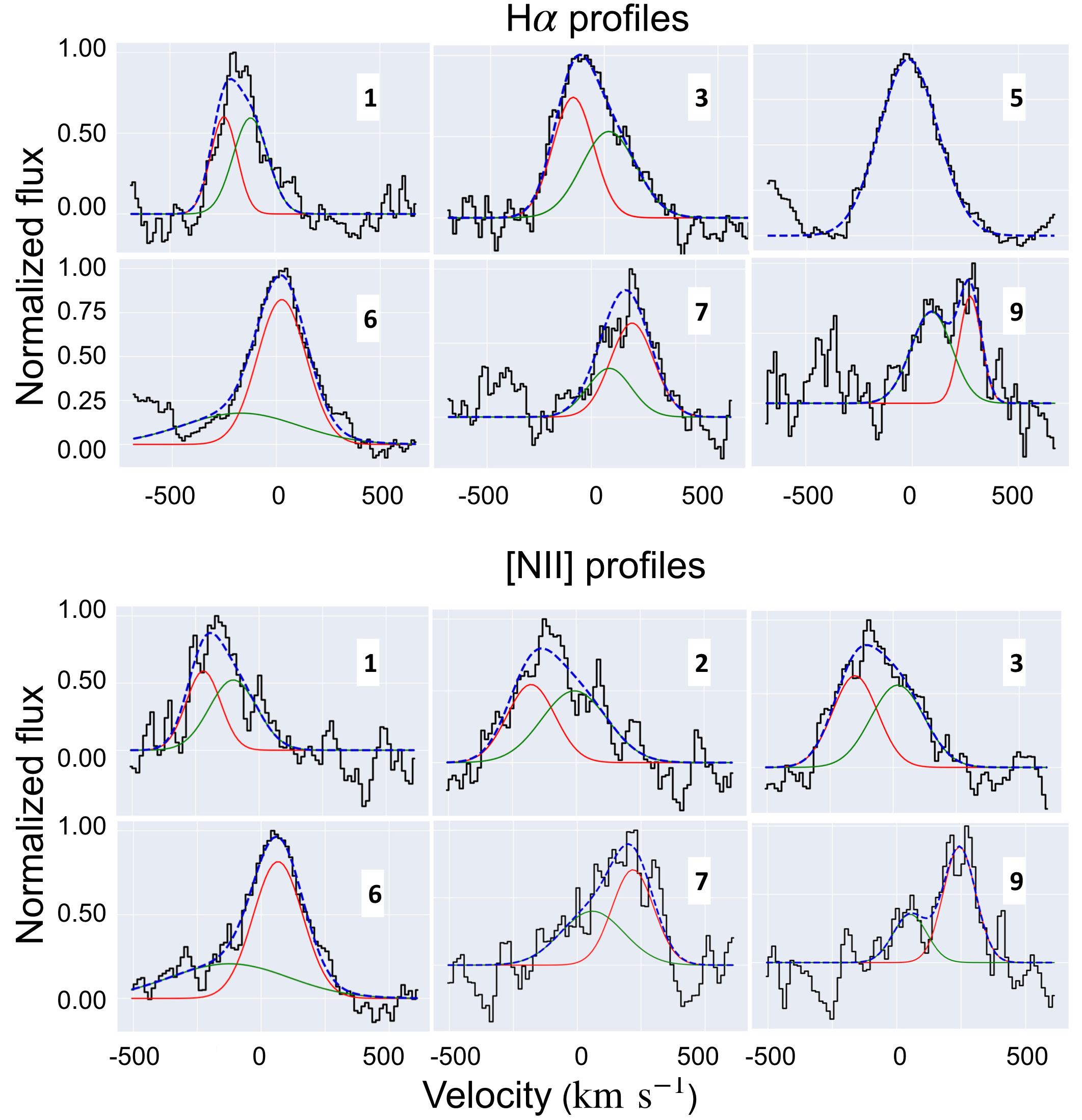}
\caption{ [\textit {Top panel}] The H$\alpha$ emission lines from ESI for slit 3 position of the first target galaxy are shown for different spaxels along the slit in different panels. A one or two component Gaussian model is used to model the emission lines similar to Fig.~\ref{fig:multigauss4}. The modelled line profile in blue shows the total model fit to the data (which is in black). [\textit {Bottom panel}] Model fit for [NII]$\rm \lambda 6584$ emission line profiles, similar to the sub-panel above.
\label{fig:multigauss2}}
\end{figure}

\begin{figure}[h!!!] 
\centering
\graphicspath{{./plots/}}
\includegraphics[width = 0.5\textwidth]{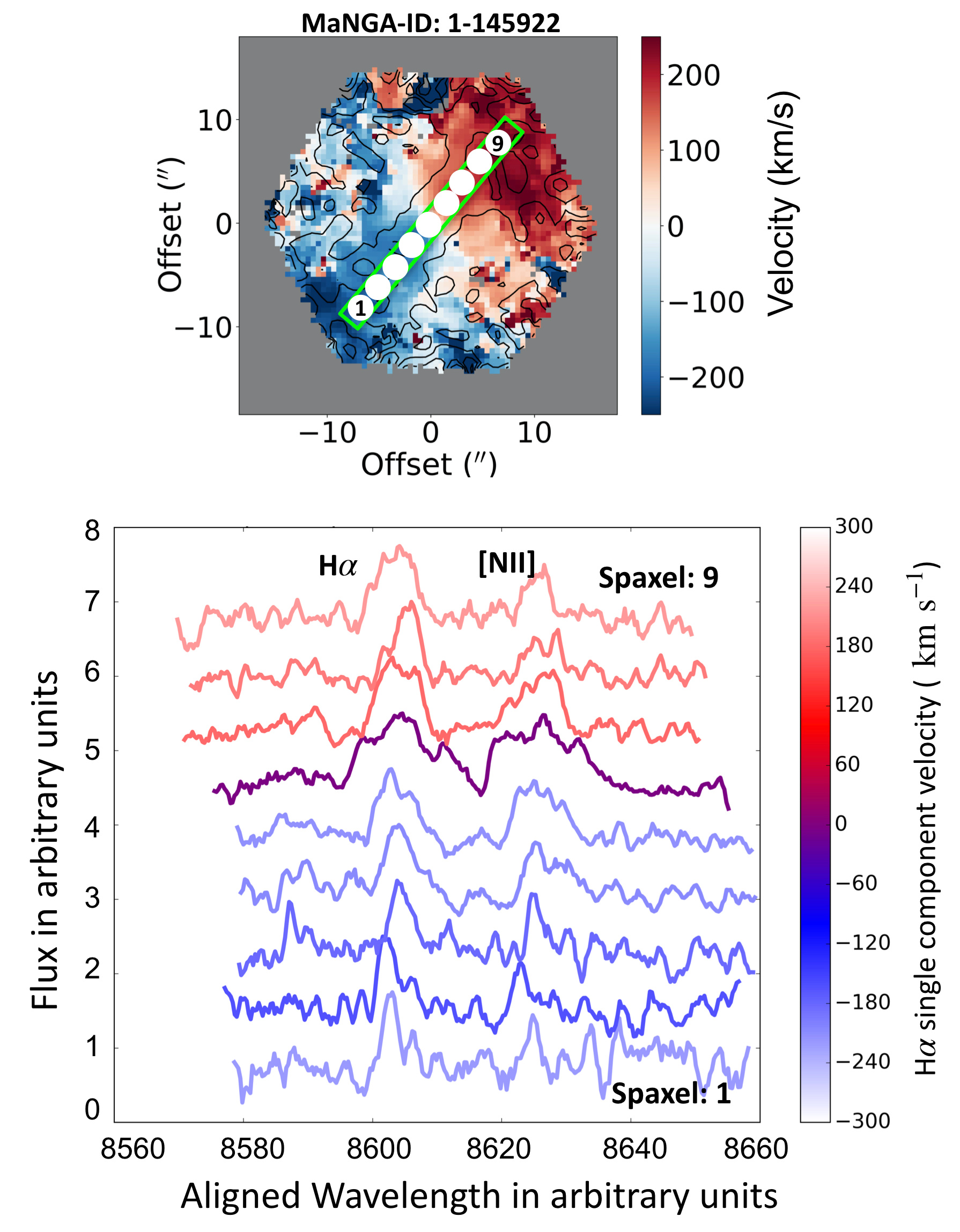}
\caption{  [\textit {Top panel}] The MaNGA ionized gas velocity field of the second target galaxy with the H$\alpha$ EW contours overplotted in black. The only slit observation with reasonable S/N taken for this galaxy is overlaid on top with different spatial bins ($\sim 1''$) indicated with circles to map the physical location of the extracted spectra on the slit. [\textit {Bottom panel}] The different rows indicate the observed ESI spectra extracted from different spaxels as marked in the top panel. The spectra are registered to the same wavelength and color coded by gas velocity. The spectra are lower signal to noise in general compared to the first target galaxy.
\label{fig:spectra}}
\end{figure}

\begin{figure}[h!!!] 
\centering
\graphicspath{{./plots/}}
\includegraphics[width = 0.5\textwidth]{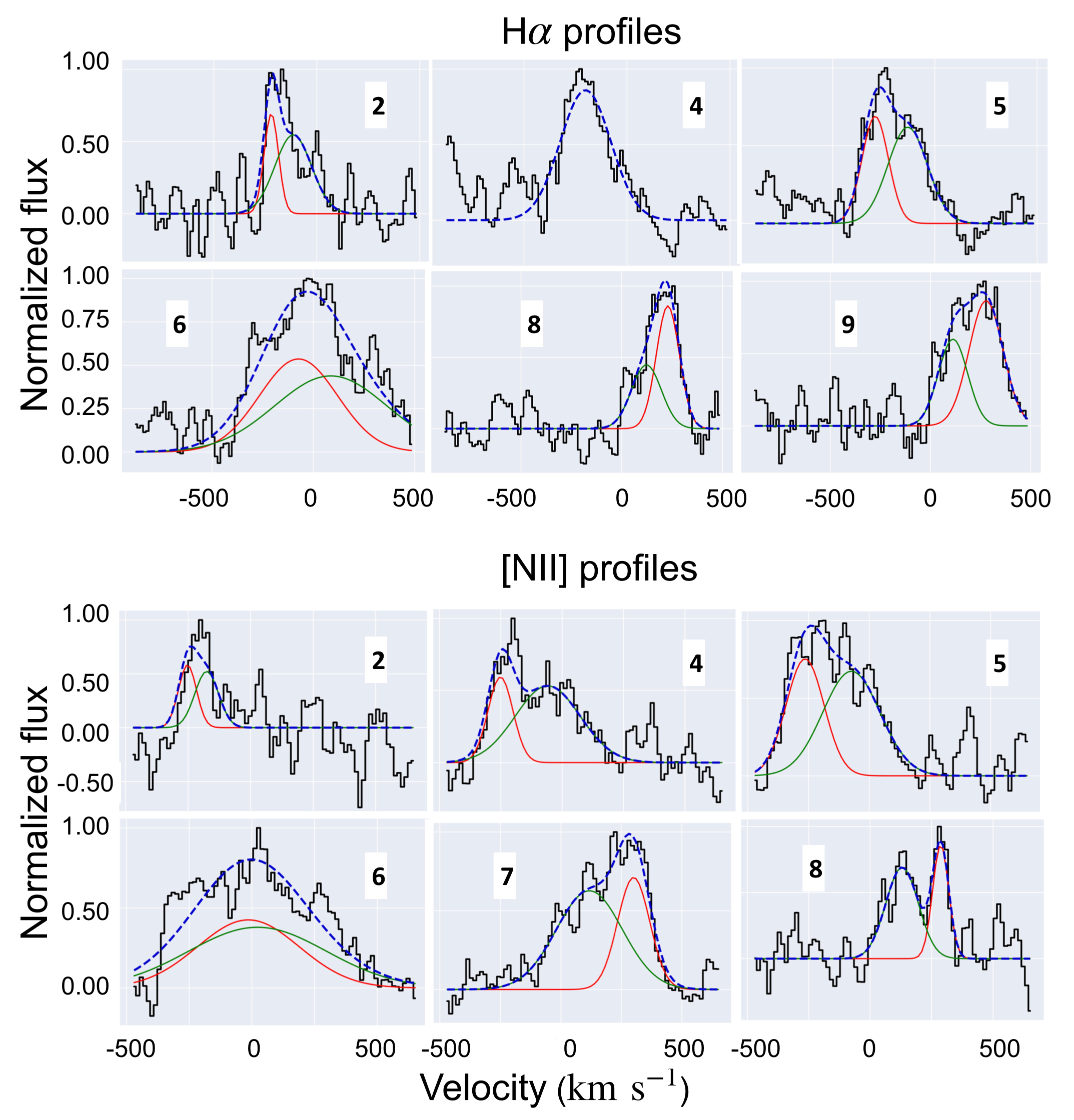}
\caption{[\textit {Top panel}] [\textit {Top panel}] The H$\alpha$ emission lines from ESI for the only slit in the second target galaxy are shown for different spaxels along the slit in different panels. A one or two component Gaussian model is used to model the emission lines similar to Fig.~\ref{fig:multigauss4}. The modelled line profile in blue shows the total model fit to the data (which is in black). [\textit {Bottom panel}] Model fit for [NII]$\rm \lambda 6584$ emission line profiles, similar to the sub-panel above.
\label{fig:multigauss}}
\end{figure}

\begin{figure*}[h!!!] 
\centering
\graphicspath{{./plots/}}
\includegraphics[width = \textwidth]{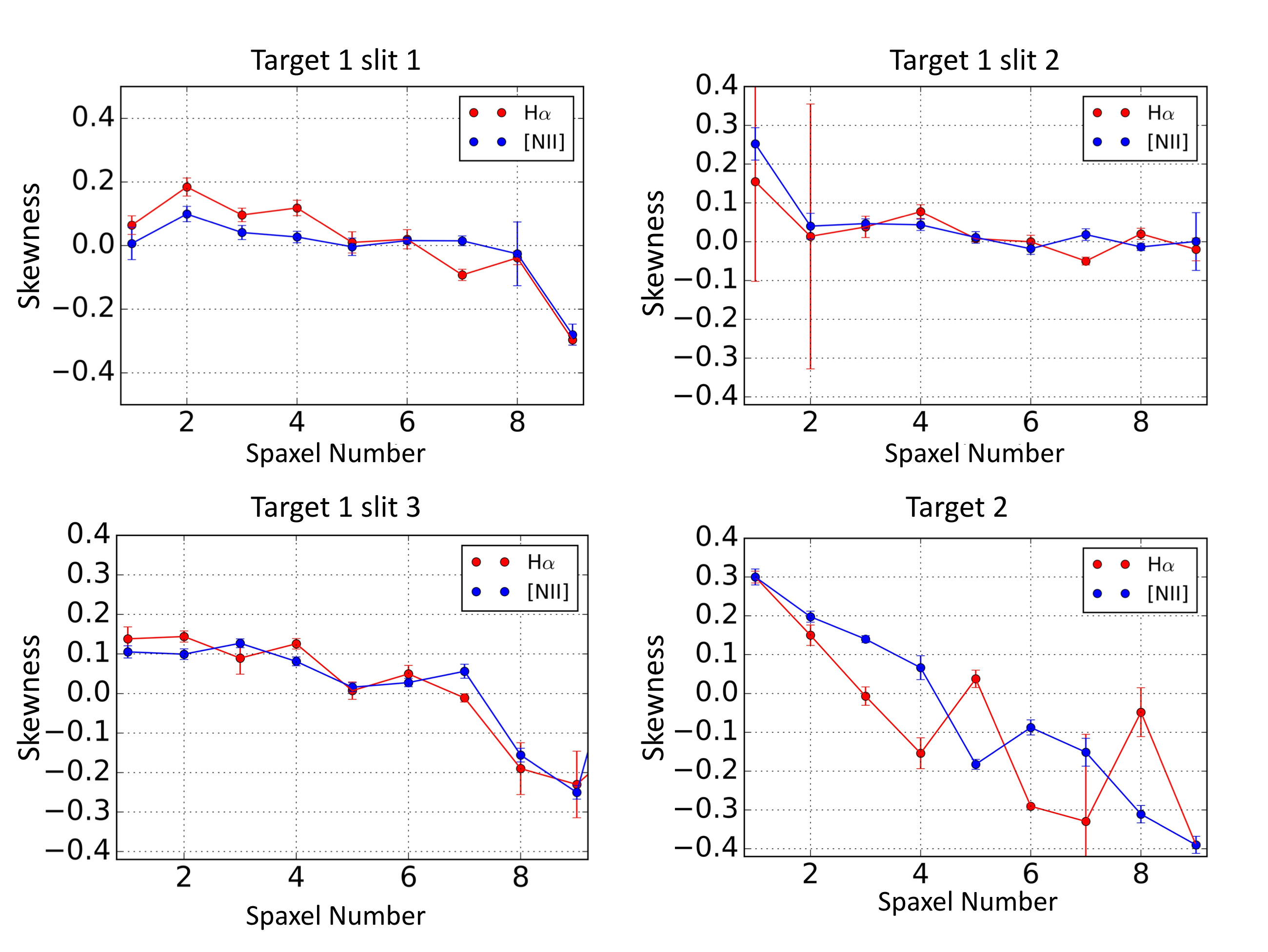}
\caption{ The observed Gauss-Hermite skewness of the H$\alpha$ (in red) and [NII] (in blue) lines for each slit positon of the two target galaxies. The skewness or the asymmetry parameter $k$, described in \S\ref{result2}, quantifies the asymmetric nature of the line profiles. $\rm k>0$ indicates a red wing and $\rm k<0$ indicates blue wing. The spaxel number maps to the spatial locations in the slit as indicated in Fig.~\ref{fig:spectra4} - \ref{fig:spectra}. Low spaxel number indicates blue-shifted part of the galaxy while high spaxel number indicates red-shifted side. 
\label{fig:skew}}
\end{figure*}

In Fig.~\ref{fig:skew}, the $k$ parameters obtained from H$\alpha$ and $\NII \lambda$6584~\AA~emission lines for all four slit observations from the two galaxies are plotted against spaxel number, which maps to locations in the slit as shown in  Fig.~\ref{fig:spectra4}$-$\ref{fig:spectra}. We can see that, for slit 1 for the first target galaxy and the only slit for the second galaxy, which sample the bi-symmetric feature in the $\rm \Ha$ EW map, the $k$ parameter values clearly transition from positive to negative as we move from the blue-shifted side (low spaxel number) towards the red-shifted side (high spaxel number) of the galaxy. This reaffirms the finding that the line profiles show a ``red'' wing on the blue-shifted side of the galaxy and a ``blue'' wing on the red-shifted side. The 2nd slit of the first galaxy shows almost a flat asymmetry parameter distribution with values oscillating close to 0. This particular slit samples the central regions of the first galaxy galaxy and reveals mostly symmetric profiles. For the 3rd slit, the $k$ parameter values again show a transition from positive to negative values similar to slit 1. This slit traces the outskirts of the galaxy parallel to the bi-symmetric feature.

\begin{figure}[h!!!] 
\centering
\graphicspath{{./plots/}}
\includegraphics[width = 0.5\textwidth]{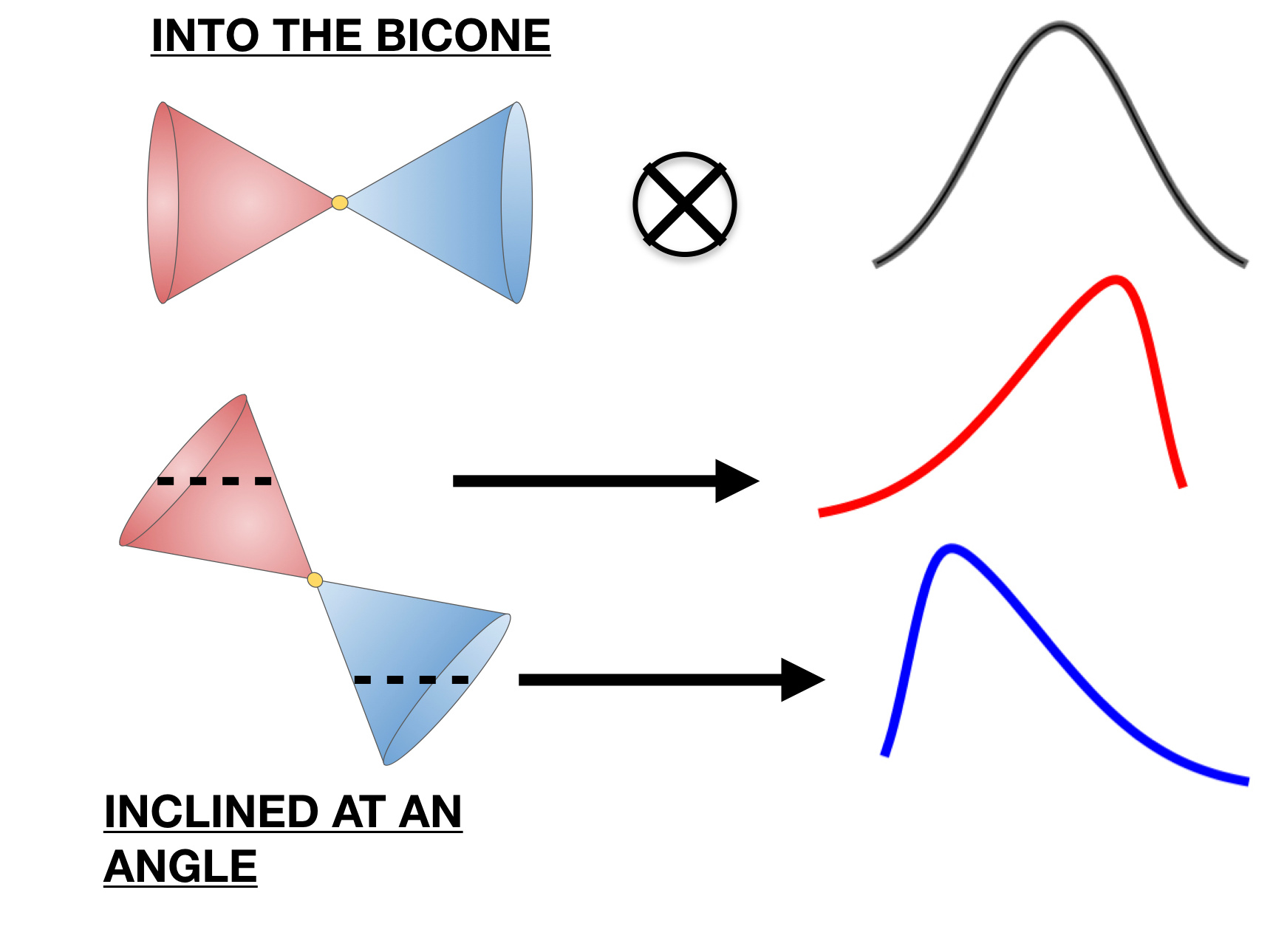}
\caption{  A schematic diagram showing how emission line profiles integrated along the line of sight of a conical wind are expected to produce asymmetric lines with a ``blue'' wing on the red side and vice versa. This behavior is evident in the emission line profiles from Keck ESI data.  
\label{fig:cartoon}}
\end{figure}





\begin{figure}[h!!!] 
\centering
\graphicspath{{./plots/}}
\includegraphics[width = 0.5\textwidth]{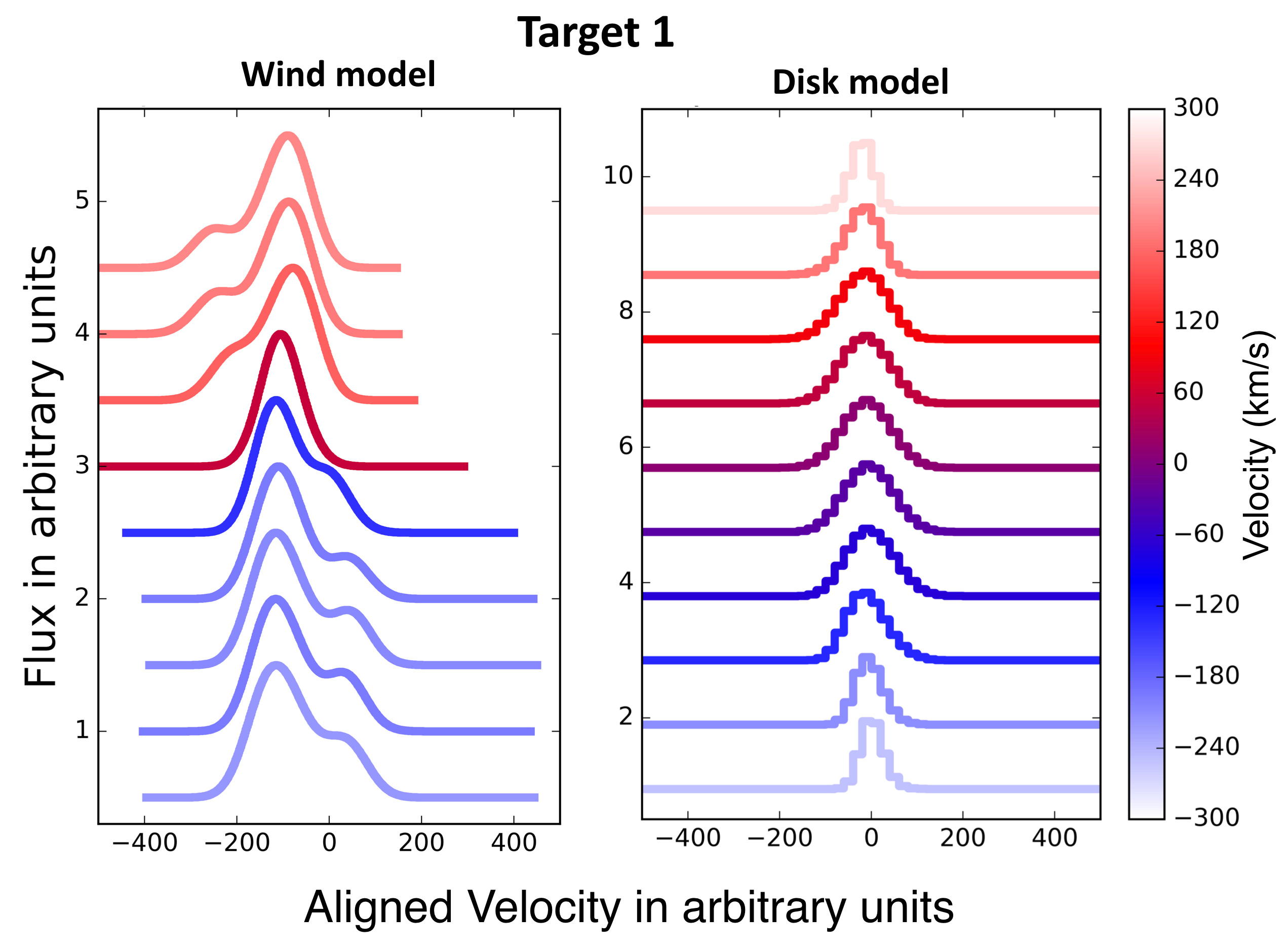}
\caption{  Left panel shows the mock spectra obtained from wind modelling of the first galaxy. Right panel shows the spectra obtained from the disk model. Each row in the plot represents the line-of-sight integrated spectra co-added inside each spatial bin sampling the same locations as ESI spaxels along the bi-symmetric feature in Fig.~\ref{fig:spectra4}. All the emission lines are registered to common velocity and the velocity information is encoded in the color scheme.  
\label{fig:winddisk1}}
\end{figure}

\begin{figure}[h!!!] 
\centering
\graphicspath{{./plots/}}
\includegraphics[width = 0.5\textwidth]{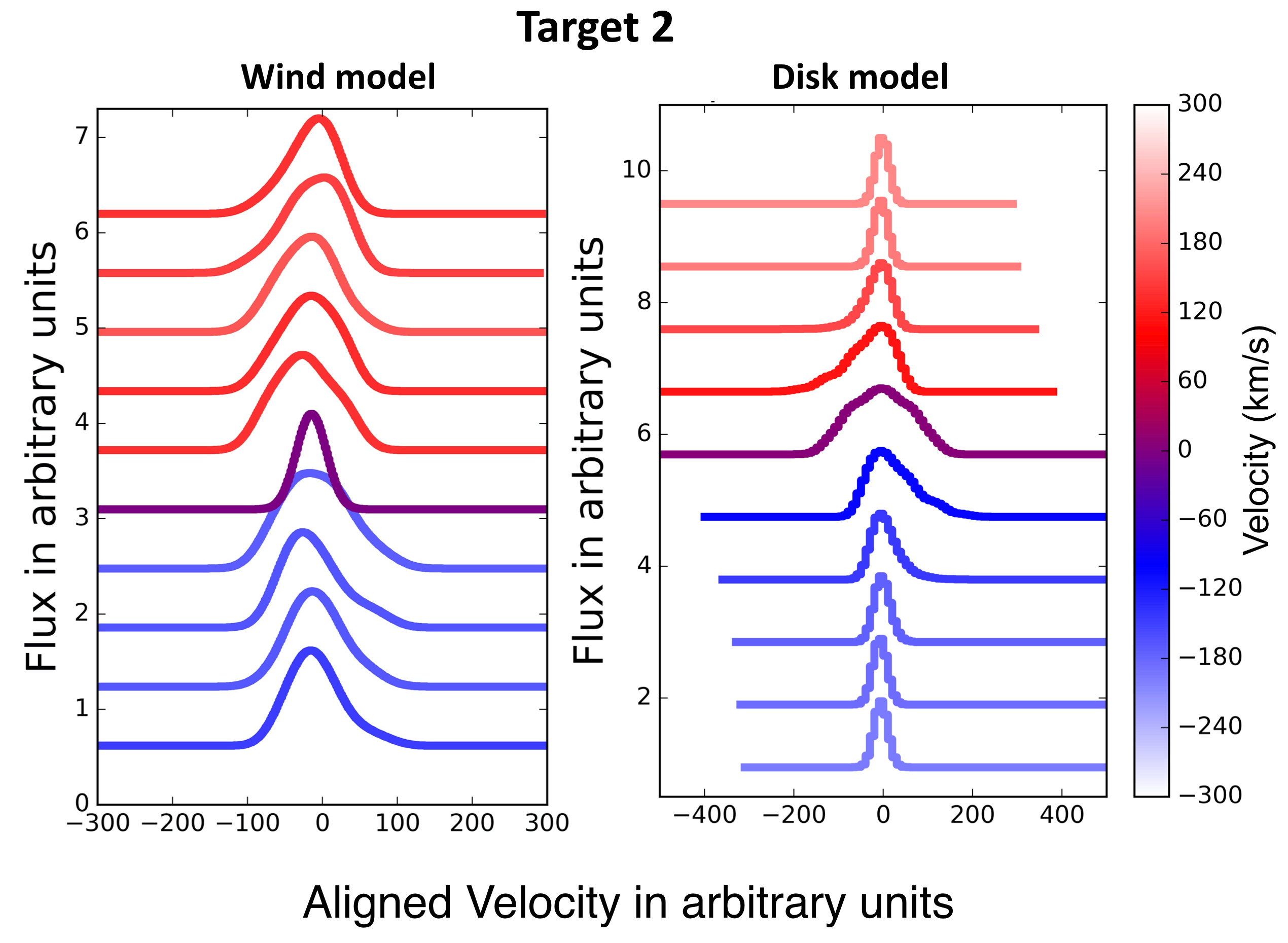}
\caption{  Left panel shows the mock spectra obtained from wind modelling of the second galaxy. Right panel shows that obtained from the disk model. Each row in the plot represents the the line-of-sight integrated spectra co-added inside each spatial bin sampling same locations as ESI spaxels along the bi-symmetric feature in Fig.~\ref{fig:spectra}. All the emission lines are registered to common velocity and the velocity information is encoded in the color scheme.
\label{fig:winddisk2}}
\end{figure}

\begin{figure}[h!!!] 
\centering
\graphicspath{{./plots/}}
\includegraphics[width = 0.48\textwidth]{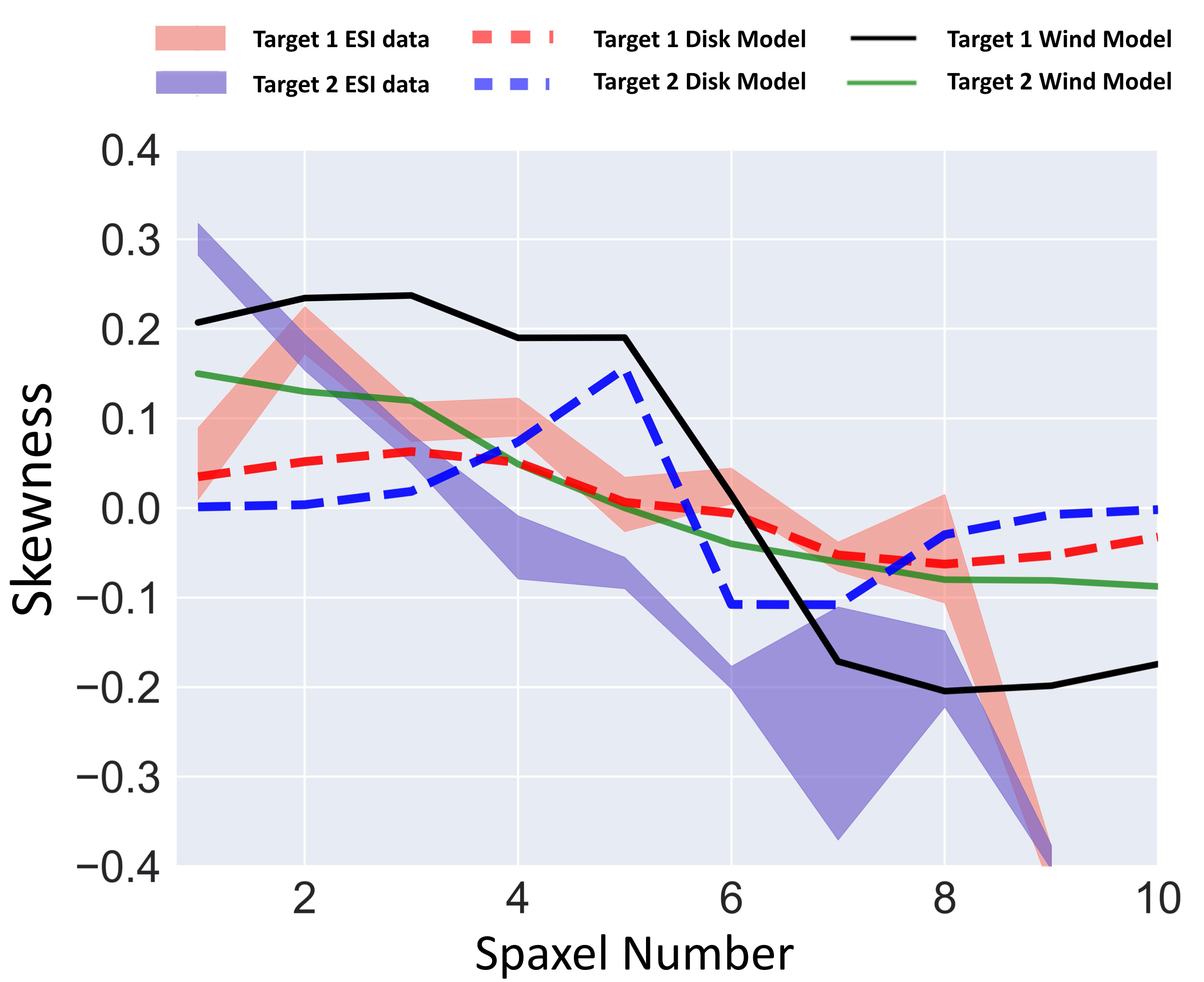}
\caption{  Comparison of the variation of asymmetry parameter obtained from ESI line profiles with predictions from different models. The salmon and blue shaded regions show the variation of the asymmetry parameter computed by averaging over $k$ values from $\Ha$ and $\NII$ lines along the presumed wind cone for both galaxies. The asymmetry parameter obtained from the disk models (shown in dashed lines) and wind models (in solid lines) for the two galaxies are overplotted on top.  
\label{fig:moddat}}
\end{figure}

\section{Discussion} \label{sec:discussion}

The main result of our observations is the systematic asymmetry of line profiles, i.e. a red wing on the blue-shifted side of the galaxy and a blue wing on the red-shifted side with symmetric profile near the center, which is observed along the bi-symmetric emission feature in both the target red geysers.
Before we discuss possible interpretations in detail, we summarize our Keck ESI data using the skewness parameter $k$ (\S\ref{result2}) measured as a function of distance from the galaxy center along the slit (Fig.~\ref{fig:skew}, also shown later in Fig.~\ref{fig:moddat}).

We will consider two physical interpretations that can give rise to our observations: ionized gas in rotation and a wide-angle outflowing wind driven by an AGN.


\subsection{Gas in rotation scenario: Disk model} \label{diskmodel}
 In the case of a rotating disk, because the portion of the disk intersected by the line-of-sight is constant due to symmetry, there is no intrinsic velocity asymmetry. Instead, asymmetry profiles mainly arise from beam smearing from the PSF and the spatial binning of the data, sometimes producing a similar red wing in the blue-shifted side of the galaxy and vice-versa, with the strength of asymmetry roughly proportional to the velocity gradient. 
Beam smearing can also inflate the FWHM of the lines especially along the regions where the projected line-of-sight velocity gradient is maximum \citep{epinat10, davis13, green14, burkert16}. 

 Although the red geysers show multiple characteristics that argue against the disk hypothesis \citep[see][]{cheung16}, it is important to address whether the Keck ESI data provides additional insight. Therefore, we construct a symmetric disk model. We use the python version of the KINMS package \citep[KINematic Molecular Simulation,][]{davis13}\footnote{https://github.com/TimothyADavis/KinMSpy} to construct the disk model. The major advantage of using KINMS is that it self-consistently accounts for beam smearing by mocking observations of a disk defined by the rotation curve and surface brightness profile. We  assume a thin disk inclined at an angle from the observer, with velocity given by the following formula:
\begin{equation*}
   \rm  V(r,\theta) = \frac{2~V_{max}}{\pi}\ tan^{-1}(r/r_{turn})
\end{equation*}

Here, V$_{\rm max}$ is the maximum velocity across the galaxy, 
and r$_{\rm turn}$ is the turnover radius for the rotation curve. 

We assign an $\Ha$ surface brightness distribution to our disk models that reproduces the MaNGA $\Ha$ flux maps for our two galaxies, as determined by fitting an exponentially declining function of the form $ \rm  \Sigma(r) \propto e^{-\frac{r}{d_{scale}}}$, where $\rm d_{scale} $ is the scale radius determined by the model fit.  
We experiment with different choices for the input parameters to find the best match between the modeled rotation curve and the position velocity curve observed from the ESI data for both the targets (see Appendix \ref{appen:velocity}). The chosen set of disk model parameters for the first galaxy are: inclination $\rm = 35^{\circ}$, turnover radius $= \rm 9''$ and maximum velocity $= \rm 715~km~s^{-1}$. The spatial scale and extent of the model grid are chosen from the ESI data. For the second galaxy, the values are: inclination $\rm = 40^{\circ}$, turnover radius $= \rm 1''$ and maximum velocity $= \rm 350~km~s^{-1}$. 
 We provide the point spread function (PSF) of FWHM $\sim 0.9 ''$, which roughly corresponds to the median seeing of Keck, as the ``beam-size'' for the computation of beam smearing. We extract the projected velocity fields and line profiles at each keck ``spaxel''. We then calculate the asymmetry parameter ($k$) of the extracted line profiles in a similar way as the data (described in \S \ref{result2}). The mock spectra extracted from the disk model along the major axis for both targets at the spatial locations of the keck spaxels are shown in Fig.~\ref{fig:winddisk1} and \ref{fig:winddisk2}.

 We find that while the disk models can produce considerable asymmetry in the line profiles because of the beam smearing effect, the nature of the resulting asymmetry does not match the data. For the first galaxy, the observed position velocity curve from ESI data is almost linear with a gentle slope (Appendix \ref{appen:velocity}, Fig.~\ref{fig:pv}). This results in the skewness parameter due to beam smearing of the disk (which approximately varies according to the line-of-sight velocity gradient) to rise and decline gently with less asymmetry overall compared to the data. 
 For the second galaxy, the observed velocity curve has a sharp gradient near the center with considerable flattening at the edges (Appendix \ref{appen:velocity}, Fig.~\ref{fig:pv}). Hence the skewness from the disk model is greater near the center compared to the outskirts.  These findings are in accordance with other similar studies of beam smearing on star forming disk galaxies \citep{green14}. Thus, the increasing asymmetry towards the outskirts of the galaxies along the kinematic axis, as observed from ESI spectra, can not be captured by the disk model. 
The trend of asymmetries that we obtain from the disk model for both galaxies along the major axis of the disk are shown by (red and blue) dashed lines in Fig. \ref{fig:moddat}. 

We see in \S\ref{target1} and Fig.~\ref{fig:multigauss2} that the third ESI slit, sampling locations offset to the bi-symmetric feature in the first galaxy, also shows considerable asymmetry in the observed ESI spectra with similar trend as locations along the bi-symmetric feature (Fig.~\ref{fig:skew}). This level of asymmetry is however not seen at similar spatial locations from the disk model, which shows a skewness value oscillating around $\sim 0$. This is because the velocity gradient declines rapidly as we move parallel to major axis of the disk away towards the outskirts. Thus this model struggles to reproduce both the trend and the magnitude of the skewness parameter for locations offset to the kinematic major axis. 

We find that the FWHM of the secondary velocity component in the observed spectra at several locations in both galaxies exceed $\rm 500-600~km~s^{-1}$, a value which is quite high and is typically attributed to ionized gas outflows in numerous studies \citep{arribas14, fischer17, humire18, cuoto17, pinto19}. The mock spectra extracted from the disk model for our target galaxies show an average enhancement of the velocity dispersion of about $\rm 40 - 50~km~s^{-1}$ (or FWHM upto $\rm 100~km~s^{-1}$) due to beam smearing effect, far below what we observe from data. Fig.~\ref{fig:winddisk1} and \ref{fig:winddisk2} (right panels) demonstrate that apart from the central spaxel that shows the FWHM value to be much higher than average, exceeding $\sim \rm 150~km~s^{-1}$, the modelled spectral lines from beam smeared disks are quite narrow in the outskirts. On the other hand, the ESI spectra for both galaxies show significantly high FWHM near the center as well as in the outskirts of the galaxies, in locations along, offset or perpendicular to the bi-symmetric feature. In other words, the enhanced dispersion doesn't necessarily follow the velocity field gradient which implies that beam smearing effect is not the primary reason behind the observed broad lines.



\subsection{Outflowing gas scenario: Wind model} \label{model}

Let us now consider the scenario of radially outflowing motion of gas particles in a filled bi-cone geometry. 
We use a simple wind model to test our interpretation of the ESI emission line velocity profiles. A 3D cartesian grid of points is populated with tracer particles with uniform density that are weighted by a Hernquist profile \citep{hernquist}. 
 The outflowing wind is assumed to be a filled wide angle bi-cone centered on the galaxy. Each gas particle in the bicone is given a constant radially outward velocity directed away from the center. Those outside the bicone are given zero velocity and zero weight. The model is motivated by our observation that the dominant ionization in the red geysers is extended LI(N)ER-like (Fig.~\ref{fig:prototype} \& \ref{fig:prototype2}) and hence the assumption that the strong emission lines (e.g., H$\rm \alpha$, [NII]) are ionized mainly by the radiation field of evolved post-AGB stars present around it \citep{yan12, belfiore16}. Although shocks can also contribute to the ionization, the smooth components of the $\Ha$ flux distribution in red geysers typically fall off roughly proportional to the stellar surface brightness supporting a stellar origin for the observed LIER ionization. Moreover, the exact nature of the ionization source should not impact the model-predicted velocity profiles to first order as both models are constrained by the observed $\Ha$ flux distribution. The warm ionized gas clouds entrained by
the wind trace the observed kinematics and emit emission line flux due to the assumed ionization. Hence the projected velocity field is a convolution of the wind geometry along the line of sight and the galaxy's 3D luminosity profile. The projected line-of-sight component of the wind velocity at each point inside the bicone is weighted by its Hernquist
profile value. In order to construct the luminosity distribution which gives rise to the wide-spread ionization of the observed gas, imaging and dynamical constraints on the stellar component,  obtained from Jeans Anisotropic Modeling \citep[JAM,][]{cappellari08} are used for both target galaxies in the model. We use JAM analyses to find intrinsic (3D) axis ratio and inclination of the galaxy, while the projected major axis effective radius and sky position angle are obtained from the Nasa Sloan Atlas (NSA) catalogue. For target 1-217022, we have taken the axis ratio to be 0.4, galaxy inclination as 50$^{\circ}$, a projected major-axis effective radius to be 7$''$, and an on-sky PA as 53$^{\circ}$ \citep[see Methods: Dynamical modeling evidence against the presence of disks, ][]{cheung16}. For the second galaxy (target 1-145922), the intrinsic axis ratio is taken to be 0.5, inclination as 20$^{\circ}$, a projected major-axis effective radius to be 9$''$ , and an on-sky PA as -73$^{\circ}$. The detailed description of the JAM parameters for the second galaxy is presented in Appendix \ref{appen:jam}. The wind parameters like the opening angle, length, intrinsic velocity and the inclination of the wind cone are varied manually until the best qualitative match is obtained between the modeled and the observed MaNGA 2D gas kinematics.
For the first target galaxy, the best match is found for wind opening angle $= \rm 80^{\circ}$, inclination $\rm = 75^{\circ}$, position angle $= 55^{\circ}$ and velocity $\rm = 300~km~s^{-1}$. For the second galaxy, the wind parametrs are opening angle $= \rm 40^{\circ}$, inclination $= \rm-55^{\circ}$, PA $= 135^{\circ}$, and velocity $\rm = 200~km~s^{-1}$.  


We construct a data cube consisting of spatial and spectral dimensions from the model. We assign single Gaussian emission line profiles to model gas particles throughout the 3D volume of the cone and convolve the flux distribution with a Gaussian of FWHM = 0.9$''$ along each spatial dimension to capture the effect of beam smearing on the simulated data cube. The velocity dispersion is taken to be the instrumental dispersion ($16~\rm km~s^{-1}$). Finally, we construct spatial bins following the same slit orientation as the Keck ESI data and integrate the line profiles along lines of sight to produce mock spectra that can be compared to the ESI data.


Due to the radially outward velocity of the gas particles inside the cone, material with the greatest line-of-sight velocity component contributes the most in observed velocity and flux. Since the velocity components are integrated along the line of sight, this leads to a series of fainter components with velocities changing slightly in a systematic manner and they add up together to give a ``winged'' emission line profile near the two ends of the cone (Fig. \ref{fig:cartoon}).

Moving closer towards the center of the galaxy, the volume inside the cone decreases. This leads to less gas to integrate along the line of sight. The velocity variation is also comparatively less because of a much tighter configuration and there is lesser variation in the angle of the velocity vector to the observer. Consequently, we expect a rather symmetric profile near the central part of the galaxies. Parallel to the axis of the cone and towards the outskirts of the galaxy, we trace the edge of the cone according to the wind hypothesis. Although the volume of the cone and the number of gas particles present towards the edge is smaller compared to that of the axis, similar asymmetries are expected though of lesser extent.


We have extracted mock spectra from the simple outflowing wind model we constructed for both our targets along the spatial positions that match the orientation of our slits and have color-coded them by their respective observed velocity (Fig.~\ref{fig:winddisk1} and \ref{fig:winddisk2}). 
 The model spectra, after properly accounting for beam smearing (as we did in the disk model), have been binned spatially to match the ESI extractions. The extracted spectra in different spatial bins are plotted in different rows after being registered to a common wavelength. As in the real data, we see a clear blue-wing in the red-shifted side and a red wing on the blue side for the first galaxy thus agreeing to the wind cone hypothesis for this galaxy.  An additional observation from the ESI slit sampling the bi-symmetric feature is that the ``blue'' wing in the red-shifted side (and vice versa on the other side) has such a big offset from the primary component that it actually lands in the blue-shifted side relative to the systematic velocity. In other words, the velocity offset between the primary and secondary component sometimes exceeds $\rm \sim 250~km~s^{-1}$ leading to primary and secondary components to be on opposite sides of the systemic velocity. This has also been reproduced by the wind model and possibly suggests that the cone opening half-angle could be larger than the inclination and it covers the plane of the sky. 

For the wind modelling of the second galaxy, although extended wings and the systematic variation across the slit are observed, the amplitude of asymmetry is found to be weaker than the first galaxy. This might be due to the fact that the velocity field obtained from the toy wind model for the second galaxy did not fit the data as well. Further, the emission signal of the second galaxy drops considerably towards the outskirts, which makes model comparison more difficult. 



 We compute the asymmetry parameter $k$ as discussed in \S\ref{result2} for the wind models and compare the $k$ profiles from these simulated spectra to the ESI data and the previously discussed disk models in Fig. \ref{fig:moddat}. The salmon and blue shaded regions show the average asymmetry parameter variation, computed by averaging the $k$ values of both $\Ha$ and $\NII \lambda$6584~\AA~ emission lines as observed from ESI data, along the bi-symmetric feature for the two target galaxies. The asymmetry parameters obtained from the disk models for those two galaxies are overplotted in dashed lines, while that from the wind model are shown in solid lines. Clearly in either targets, neither the rotating disk nor outflowing wind models fully reproduce the observed asymmetry profiles. Both models are able to reach typical values we observe for $k$ ($\rm\pm 0.2-0.25$), but the disk models only achieve these asymmetries near the center, where beam smearing of a strong velocity gradient reaches a maximum.  In the galaxy outskirts, the disk model asymmetries nearly vanish while the observed k magnitudes continue to increase. Both wind models reproduce this increased asymmetry in the outskirts, although they fail to match the observed magnitude of $k$, especially for Target 2. This might be explained generally in the wind interpretation by a clumpy or turbulent outflowing medium.  It is harder to explain with a rotating disk which requires a degree of dynamical stability in order to maintain its apparent velocity structure.



%


 %


\section{Conclusion} \label{sec:conclusion}

We have performed an analysis of the  emission line velocity profiles of H$\alpha$ and [NII] $\lambda$6584 of two red geyser galaxies using high spectral resolution (R$\sim$8000) Keck-ESI  observations. Our observations  of the first target (MaNGAID: 1-217022) include three slit orientations that sample different parts of the galaxy, namely, the bi-symmetric feature, the regions around the center and the outskirts. The second target (MaNGAID: 1-145922) includes only one slit observation along its bi-symmetric feature. The slits that align with the bi-conical axis for both galaxies show strong asymmetry in the emission lines. The slit that lie parallel to the bi-symmetric feature with an offset for the first galaxy also shows similar asymmetry. The shape of the emission lines, which can be decomposed into primary and secondary velocity components, exhibit a red wing on the blue-shifted part of the galaxy and a blue wing on the red side, with a symmetric profile near the center. This has been quantified by an asymmetry parameter that changes systematically from positive to negative values (Fig.~\ref{fig:skew} and \ref{fig:moddat}). A bi-cone geometry of the gas with a radially outward motion can better explain the observed features than a rotating gas disk.

The presence of low-luminosity radio mode AGNs in the red geysers \citep{roy18} along with the confirmation of an outflowing wind scenario obtained from MaNGA data and the current observations from Keck ESI, lead to the evidence of AGN-driven winds in the red geyser galaxies. 

\section*{Acknowledgement} 
This research was supported by the National Science Foundation under Award No. 1816388. The authors thank the anonymous referee for helpful suggestions that significantly improved the manuscript. NR thanks Professor Timothy Davis for assisting with the implementation of the KINMSpy package. NR thanks Professor Puragra Guhathakurta and Jason Xavier Prochaska for helpful comments and discussions. SC acknowledges support from the Science and Technology Funding Council (STFC) via a PhD studentship (grantnumber ST/T506448/1).
RAR thanks partial financial support from Conselho Nacional de Desenvolvimento Cient\'ifico e Tecnol\'ogico (202582/2018-3 and 302280/2019-7) and Funda\c c\~ao de Amparo \`a pesquisa do Estado do Rio Grande do Sul (17/2551-0001144-9 and 16/2551-0000251-7). The authors also wish to acknowledge the very significant cultural role and reverence that the summit of Mauna Kea has always had within the indigenous Hawaiian community. We are very fortunate to have the opportunity to conduct observations from this mountain. 
Funding for the Sloan Digital Sky Survey IV has been provided by the Alfred P. Sloan Foundation, the U.S. Department of Energy Office of Science, and the Participating Institutions. SDSS-IV acknowledges
support and resources from the Center for High-Performance Computing at
the University of Utah. The SDSS web site is \href{http://www.sdss.org}{www.sdss.org}.

SDSS-IV is managed by the Astrophysical Research Consortium for the 
Participating Institutions of the SDSS Collaboration including the 
Brazilian Participation Group, the Carnegie Institution for Science, 
Carnegie Mellon University, the Chilean Participation Group, the French Participation Group, Harvard-Smithsonian Center for Astrophysics, 
Instituto de Astrof\'isica de Canarias, The Johns Hopkins University, 
Kavli Institute for the Physics and Mathematics of the Universe (IPMU) / 
University of Tokyo, the Korean Participation Group, Lawrence Berkeley National Laboratory, 
Leibniz Institut f\"ur Astrophysik Potsdam (AIP),  
Max-Planck-Institut f\"ur Astronomie (MPIA Heidelberg), 
Max-Planck-Institut f\"ur Astrophysik (MPA Garching), 
Max-Planck-Institut f\"ur Extraterrestrische Physik (MPE), 
National Astronomical Observatories of China, New Mexico State University, 
New York University, University of Notre Dame, 
Observat\'ario Nacional / MCTI, The Ohio State University, 
Pennsylvania State University, Shanghai Astronomical Observatory, 
United Kingdom Participation Group,
Universidad Nacional Aut\'onoma de M\'exico, University of Arizona, 
University of Colorado Boulder, University of Oxford, University of Portsmouth, 
University of Utah, University of Virginia, University of Washington, University of Wisconsin, 
Vanderbilt University, and Yale University.\\

\section*{APPENDIX}

\subsection{Recovering the inclination using Jeans Anisotropic Modeling (JAM) }  \label{appen:jam}

In order to construct the wind model for the second target galaxy, we recover the inclination using Jeans Anisotropic Modelling (JAM) method, detailed in \cite{cappellari08} and implemented in the \texttt{jampy} package. This method approximates the stellar potential from the surface brightness distribution using Multi-Gaussian Expansion (MGE). In this instance, our observations of the surface brightness distribution comes from R-band DESI Legacy Survey imaging \cite{Legacy} as this survey is deeper than the SDSS. We utilise the Python implementation of the method described in \cite{Cappellari2002}, \texttt{mgefit}. When constructing the MGE approximation, the flattest gaussian in the best fit is found to be overly restrictive in the JAM modelling process, as it defines the minimum inclination which can be modelled. This is inherent to the JAM method and doesn't reflect a physical limitation on the inclination. We minimise this restriction while retaining suitable accuracy using the method in \cite{Scott}. Here a limit is placed on the minimum axial ratio allowed in the MGE fit, and this limit is increased until the mean absolute deviation of the fit increases by 10\%. Using this method we found the parameters which best fit the image, whilst maximising the inclination range available. The parameters are presented in Table \ref{tab:MGE_param}.

\begin{table}
	\centering
	\caption{The parameters of the best fit MGE. Total Counts refer to the counts under each component, $\sigma$ is the width of each component in pixels, and $q_{obs}$ is the projected axial ratio of each component.}
	\label{tab:MGE_param}
	\begin{tabular}{lccr} 
		\hline
		Total Counts & $\sigma$ & $q_{obs}$ \\
		\hline
		272.886 & 2.47831 & 0.95 \\
		432.696 & 10.7368 & 0.95 \\
		165.534 & 21.1834 & 0.95 \\
		476.804 & 23.2103 & 1.0 \\
		1109.95 & 58.3299 & 1.0 \\
		\hline
	\end{tabular}
\end{table}

We perform an MCMC simulation to find the best fit parameters for the JAM model to match the $v_{RMS}$ map from the MaNGA data. We decide that a simple JAM model neglecting dark matter is adequate for measuring the inclination, as in \cite{degeneracy}. The free parameters which are allowed to vary are the orbital anisotropy $\beta$, the inclination $\cos{i}$, and the dynamical mass to light ratio $M/L$. Both $\beta$ and $M/L$ are assumed to be constant across the galaxy. Note that $M/L$ contains components from the stellar mass-to-light ratio and the presence of dark matter, and the contribution of each of these cannot be determined by this model. We use uniform priors on each of the parameters within the bounds below:
\begin{itemize}
    \item $\beta$ is allowed to vary between 0 and 0.75. Restricting $\beta > 0$ is required to break the known degeneracy between anisotropy and inclination, and is observationally motivated for fast rotators \citep[see Section 3.1.1 of ][]{degeneracy} as the red geyser population are found to be. 
    \item $\cos{i}$ is allowed to vary between 0 and $\cos{i_{min}}$, which is the minimum inclination imposed by the flattest component in the MGE approximation
    \item $M/L$ is allowed between 0 and 10
    \item $\log{f}$ is an extra parameter added to quantify any underestimation of the errors, and is assumed to have a uniform prior within range -5 and 1
\end{itemize}

\begin{figure}
	\includegraphics[width=\columnwidth]{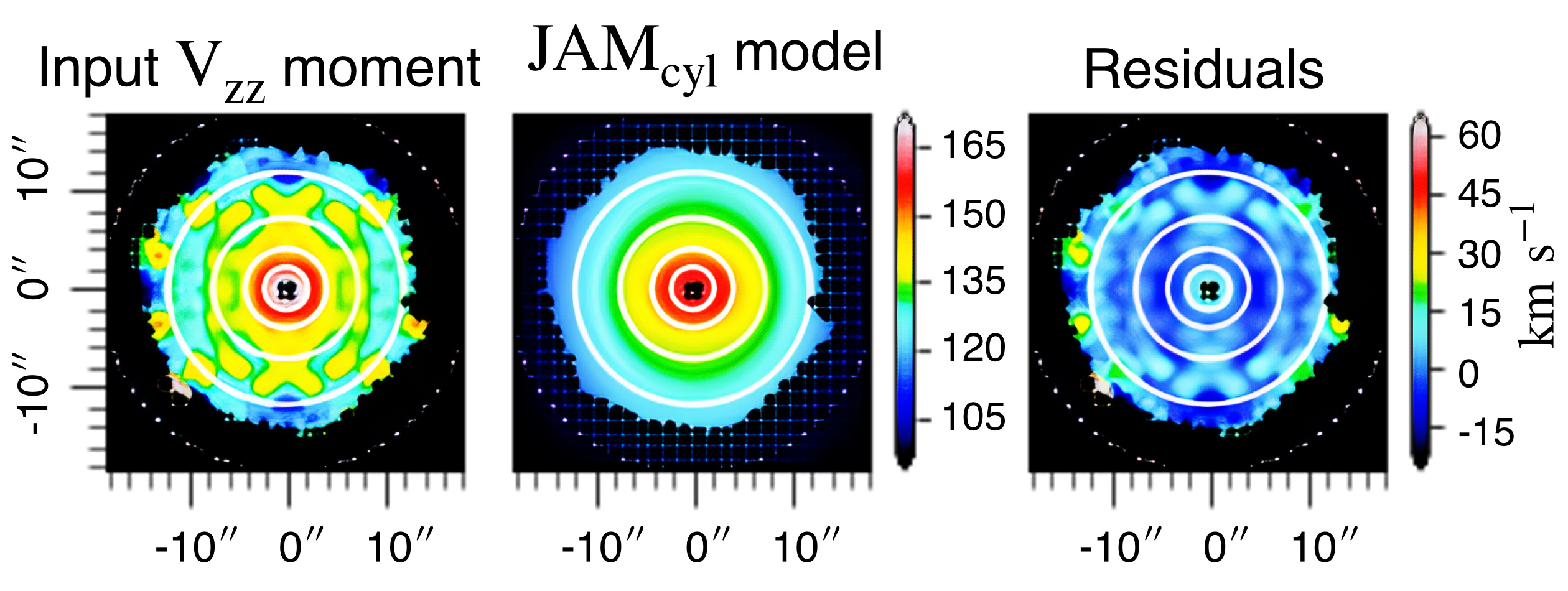}
    \caption{Left: The symmetrised $v_{rms}$ field for the galaxy which is input to JAM, Centre: The output from the best fit JAM model, Right: The residuals of the model and data. The white contours on all trace the surface brightness of the MGE. Black dots denote masked pixels. The unit of the color bars shown is $\rm km~s^{-1}$. }
    \label{fig:best_mod}
\end{figure}

We find our best fit model with $\chi ^2 /{DoF} = 2.68$. The predicted $v_{rms}$ ($v_{zz}$) from the best model is shown in comparison with the observed $v_{rms}$ in Figure \ref{fig:best_mod}, alongside the residuals. The measured parameters are presented in Table \ref{tab:JAM_param}. It is noted that the best fit inclination is close to the boundary on the prior, with artificially small errors. As such, it is believed that the true inclination may lie below the limit imposed by the MGE axial ratio. 

\begin{table}
	\centering
	\caption{The parameters of the best fit JAM model. The inclination should be interpreted as an upper limit.}
	\label{tab:JAM_param}
	\begin{tabular}{lccr} 
		\hline
		Inclination ($^{\circ}$) & $\beta$ & $M/L$ \\
		\hline
		$19.58_{-0.057}^{+0.070} $ & $0.216_{-0.029}^{+0.025}$ & $6.921_{-0.021}^{+0.019}$ \\
		
		\hline
	\end{tabular}
\end{table}

Therefore, we use an inclination of 20$^\circ$ to construct our wind model. The intrinsic 3D axis ratio is estimated from the inclination and observed axis ratio from Equation 1, \cite{weijmans14}. 

\subsection{Parameters of individual velocity components of $\Ha$ and $\NII$ emission lines}  \label{appen:fitparam}

The best fit parameters obtained from fitting the $\Ha$ and $\NII \lambda6584$~\AA~emission lines from ESI at different slit positions for both targets are shown in the following tables. The different rows in each table indicate different Keck spaxels (or spatial locations) along each slit position. If the double gaussian model is preferred for an emission line in a particular spaxel, V$_1$ and V$_2$ gives the primary and secondary velocity components while $\sigma_1$ and $\sigma_2$ are the corresponding dispersions. If the single component model is preferred, the velocities and dispersions are quoted as V$_1$ and $\sigma_1$ respectively, while keeping the other two columns empty.  

\begin{table*}[htp]
    \begin{center}
    \label{table:1}
    \caption{Best fit model parameters for H$\alpha$ and [NII] emission lines from ESI for slit 1 position of the first target galaxy.  }
     \begin{tabular}{||c| c c c c||} 
         
         \hline
         Spaxel & $\rm V_1 \pm \Delta V_1~(\rm km~s^{-1})$ & $\rm V_2 \pm \Delta V_2~(\rm km~s^{-1})$ & $\rm \sigma_1 \pm \Delta\sigma_1~(\rm km~s^{-1})$ & $\rm \sigma_2 \pm \Delta\sigma_2~(\rm km~s^{-1})$ \\ [0.5ex]  
          \hline
         
         \hline\hline
         Spaxel 1 (H$\alpha$) & -242.99$\pm$0.54 & -129.28$\pm$4.06 & 50.71$\pm$3.89 & 72.88$\pm$0.54 \\ 
         \hline
         Spaxel 1 (NII) & -232.49$\pm$1.19 & -129.92$\pm$5.01 & 55.02$\pm$4.92 & 59.97$\pm$1.19 \\ 
          \hline
         
         \hline\hline
         Spaxel 2 (H$\alpha$) & -187.46$\pm$1.59 & -44.06$\pm$10.59 & 75.09$\pm$3.29 & 129.57$\pm$1.58 \\ 
         \hline
         Spaxel 2 (NII) & -218.99$\pm$1.72 & -48.57$\pm$1.00 & 83.84$\pm$6.84 & 131.94$\pm$1.72 \\ 
          \hline
         
         \hline\hline
         Spaxel 3 (H$\alpha$) & -146.59$\pm$2.96 & -17.67$\pm$1.86 & 96.97$\pm$6 & 140.33$\pm$2.96 \\ 
         \hline
         Spaxel 3 (NII) & -176.29$\pm$1.55 & -27.29$\pm$7.10 & 87.67$\pm$6.43 & 110.11$\pm$1.55 \\ 
          \hline
         
         \hline\hline
         Spaxel 4 (H$\alpha$) & -132.44$\pm$2.73 & 20.27$\pm$1.23 & 84.10$\pm$2.95 & 113.93$\pm$2.73 \\ 
         \hline
         Spaxel 4 (NII) & -105.67$\pm$1.27 & 4.50$\pm$14.96 & 104.90$\pm$4.83 & 143.41$\pm$1.27 \\ 
          \hline
         
         \hline\hline
         Spaxel 5 (H$\alpha$) & -43.41$\pm$3.50 & ---- & 117.20$\pm$3.18 & ---- \\ 
         \hline
         Spaxel 5 (NII) & -32.20$\pm$3.10 & ---- & 132,10$\pm$2.31 & ---- \\ 
          \hline
         
         \hline\hline
         Spaxel 6 (H$\alpha$) & -6.93$\pm$0.39 & ---- & 124.28$\pm$5.16 & ---- \\ 
         \hline
         Spaxel 6 (NII) & -25.25$\pm$1.76 & 13.10$\pm$5.77 & 102.63$\pm$5.23 & 164.10$\pm$1.76 \\ 
          \hline
         
         \hline\hline
         Spaxel 7 (H$\alpha$) & 62.34$\pm$4.34 & -126.16$\pm$13.82 & 108.73$\pm$4.93 & 276.80$\pm$4.34 \\ 
         \hline
         Spaxel 7 (NII) & 31.20$\pm$1.70 & 82.55$\pm$17.48 & 88.76$\pm$6.66 & 219.95$\pm$1.70 \\ 
          \hline
         
         \hline\hline
         Spaxel 8 (H$\alpha$) & 108.87$\pm$5.49 & 92.73$\pm$3.50 & 93.15$\pm$4.87 & 278.01$\pm$5.49 \\ 
         \hline
         Spaxel 8 (NII) & 114.40$\pm$7.42 & 119.99$\pm$22.49 & 80.37$\pm$1.34 & 253.94$\pm$7.42 \\ 
          \hline
         
         \hline\hline
         Spaxel 9 (H$\alpha$) & 140.16$\pm$3.94 & -51.87$\pm$9.35 & 89.37$\pm$4.56 & 277.89$\pm$3.95 \\ 
         \hline
         Spaxel 9 (NII) & 61.46$\pm$8.06 & 232.90$\pm$8.42 & 79.12$\pm$5.09 & 79.78$\pm$8.06 \\ 
          \hline
         
         \hline\hline
    \end{tabular}
    \end{center}
\end{table*}

\begin{table*}[htp]
\begin{center}
\label{table:2}
\caption{ Best fit model parameters for H$\alpha$ and [NII] emission lines from ESI for slit 2 position of the first target galaxy.  }
 \begin{tabular}{||c| c c c c||} 
 
 \hline
 Spaxel & $\rm V_1 \pm \Delta V_1~(\rm km~s^{-1})$ & $\rm V_2 \pm \Delta V_2~(\rm km~s^{-1})$ & $\rm \sigma_1 \pm \Delta\sigma_1~(\rm km~s^{-1})$ & $\rm \sigma_2 \pm \Delta\sigma_2~(\rm km~s^{-1})$ \\ [0.5ex]  
  \hline
 
 \hline\hline
 Spaxel 1 (H$\alpha$) &  -173.12$\pm$1.24 & 21.84$\pm$4.17 & 34.40$\pm$4.31 & 103.68$\pm$1.24 \\ 
 \hline
 Spaxel 1 (NII) & -136.19$\pm$3.80 & 54.99$\pm$8.83 & 159.15$\pm$3.59 & 184.52$\pm$3.80 \\ 
  \hline
 
 \hline\hline
 Spaxel 2 (H$\alpha$) & 48.02$\pm$6.07 & 230.44$\pm$8.67 & 95.61$\pm$7.73 & 38.40$\pm$6.07 \\ 
 \hline
 Spaxel 2 (NII) & 41.76$\pm$1.88 & -39.63$\pm$4.73 & 135.75$\pm$9.72 & 102.11$\pm$18.83 \\ 
  \hline
 
 \hline\hline
 Spaxel 3 (H$\alpha$) & 111.79$\pm$16.51 & 92.17$\pm$7.53 & 92.12$\pm$8.44 & 79.78$\pm$1.65 \\ 
 \hline
 Spaxel 3 (NII) & -42.28$\pm$9.21 & 146.24$\pm$11.89 & 100.06$\pm$7.51 & 112.64$\pm$9.21 \\ 
  \hline
 
 \hline\hline
 Spaxel 4 (H$\alpha$) & 45.41$\pm$1.05 & ---- & 123.75$\pm$4.95 & ---- \\ 
 \hline
 Spaxel 4 (NII) & -37.18$\pm$1.04 & 82.60$\pm$3.92 & 130.70$\pm$4.61 & 137.55$\pm$10.40 \\ 
  \hline
 
 \hline\hline
 Spaxel 5 (H$\alpha$) & -8.29$\pm$1.96 & ---- & 1.3967$\pm$2.11 & ---- \\ 
 \hline
 Spaxel 5 (NII) & -1.71$\pm$0.53 & ---- & 145.40$\pm$8.79 & ---- \\ 
  \hline
 
 \hline\hline
 Spaxel 6 (H$\alpha$) & -41.84$\pm$1.65 & ---- & 151.02$\pm$7.35 & ---- \\ 
 \hline
 Spaxel 6 (NII) & -33.84$\pm$1.50 & ---- & 140.20$\pm$7.27 & ---- \\ 
  \hline
 
 \hline\hline
 Spaxel 7 (H$\alpha$) &  -64.86$\pm$8.12 & ---- & 146.77$\pm$6.73 & ---- \\ 
 \hline
 Spaxel 7 (NII) & -41.25$\pm$4.55 & ---- & 139.69$\pm$3.81 & ---- \\ 
  \hline
 
 \hline\hline
 Spaxel 8 (H$\alpha$) & 39.96$\pm$4.77 & -151.19$\pm$9.21 & 51.80$\pm$4.09 & 271.33$\pm$4.77 \\ 
 \hline
 Spaxel 8 (NII) & -137.07$\pm$4.43 & 52.81$\pm$9.85 & 1.95.55$\pm$9.33 & 115.53$\pm$4.43 \\ 
  \hline
 
 \hline\hline
 Spaxel 9 (H$\alpha$) & 46.26$\pm$3.46 & 189.63$\pm$11.97 & 127.02$\pm$19.58 & 50.51$\pm$14.66 \\ 
 \hline
 Spaxel 9 (NII) & 1.18$\pm$3.80 & -136.73$\pm$67.85 & 43.03$\pm$10.15 & 263.34$\pm$3.80 \\ 
  \hline
 
 \hline\hline
\end{tabular}
\end{center}
\end{table*}

\begin{table*}[htp]
\begin{center}
\label{table:3}
\caption{ Best fit model parameters for H$\alpha$ and [NII] emission lines from ESI for slit 3 position of the first target galaxy.  }
 \begin{tabular}{||c| c c c c||} 
 
 \hline
 Spaxel & $\rm V_1 \pm \Delta V_1~(\rm km~s^{-1})$ & $\rm V_2 \pm \Delta V_2~(\rm km~s^{-1})$ & $\rm \sigma_1 \pm \Delta\sigma_1~(\rm km~s^{-1})$ & $\rm \sigma_2 \pm \Delta\sigma_2~(\rm km~s^{-1})$ \\ [0.5ex]  
 \hline\hline
 Spaxel 1 (H$\alpha$) & -242.78$\pm$0.33 & 115.84$\pm$3.80 & 63.00$\pm$3.20 & 82.60$\pm$0.33 \\ 
 \hline
 Spaxel 1 (NII) & -219.22$\pm$2.02 & -100.54$\pm$10.67 & 67.73$\pm$7.41 & 94.44$\pm$2.02 \\ 
  \hline
 
 \hline\hline
 Spaxel 2 (H$\alpha$) & -228.87$\pm$0.59 & -60.39$\pm$4.79 & 68.89$\pm$5.45 & 90.11$\pm$0.59 \\ 
 \hline
 Spaxel 2 (NII) & 176.52$\pm$1.70 & 5.75$\pm$9.39 & 93.44$\pm$5.57 & 128.30$\pm$1.71 \\ 
  \hline
 
 \hline\hline
 Spaxel 3 (H$\alpha$) & -136.26$\pm$5.24 & 20.39$\pm$1.59 & 89.23$\pm$3.99 & 119.42$\pm$5.24 \\ 
 \hline
 Spaxel 3 (NII) & 151.17$\pm$3.78 & 17.82$\pm$7.20 & 89.01$\pm$3.94 & 105.79$\pm$3.78 \\ 
  \hline
 
 \hline\hline
 Spaxel 4 (H$\alpha$) & -41.16$\pm$0.38 & ---- & 139.24$\pm$3.26 & ---- \\ 
 \hline
 Spaxel 4 (NII) & -47.89$\pm$5.90 & 143.07$\pm$15.01 & 116.24$\pm$3.60 & 80.71$\pm$5.90 \\ 
  \hline
 
 \hline\hline
 Spaxel 5 (H$\alpha$) & -19.98$\pm$2.74 & ---- & 138.34$\pm$2.19 & ---- \\ 
 \hline
 Spaxel 5 (NII) & -57.41$\pm$5.27 & 48.12$\pm$1.57 & 119.26$\pm$2.90 & 133.11$\pm$5.27 \\ 
  \hline
 
 \hline\hline
 Spaxel 6 (H$\alpha$) & 26.92$\pm$1.33 & -171.21$\pm$2.21 & 117.79$\pm$1.62 & 279.54$\pm$1.33 \\ 
 \hline
 Spaxel 6 (NII) & -9.25$\pm$0.74 & 124.43$\pm$4.54 & 114.98$\pm$3.48 & 127.23$\pm$7.47 \\ 
  \hline
 
 \hline\hline
 Spaxel 7 (H$\alpha$) & 77.83$\pm$1.80 & 117.47$\pm$5.88 & 107.12$\pm$2.29 & 279.48$\pm$1.87 \\ 
 \hline
 Spaxel 7 (NII) & 66.36$\pm$2.86 & -125.50$\pm$8.55 & 97.00$\pm$2.82 & 232.33$\pm$2.86 \\ 
  \hline
 
 \hline\hline
 Spaxel 8 (H$\alpha$) & 150.62$\pm$4.20 & 89.54$\pm$8.43 & 119.33$\pm$15.36 & 69.79$\pm$4.20 \\ 
 \hline
 Spaxel 8 (NII) & 90.88$\pm$17.56 & 96.00$\pm$9.07 & 113.29$\pm$9.74 & 81.11$\pm$17.56 \\ 
  \hline
 
 \hline\hline
 Spaxel 9 (H$\alpha$) & 195.31$\pm$4.15 & 86.16$\pm$11.18 & 104.25$\pm$18.22 & 105.42$\pm$4.15 \\ 
 \hline
 Spaxel 9 (NII) & 217.90$\pm$11.40 & 55.42$\pm$2.74 & 81.27$\pm$6.90 & 121.58$\pm$11.40 \\ 
  \hline
 
 \hline\hline
\end{tabular}
\end{center}
\end{table*}

\begin{table*}[htp]
\begin{center}
\label{table:4}
\caption{ Best fit model parameters for H$\alpha$ and [NII] emission lines from ESI for the second target galaxy.  }
 \begin{tabular}{||c| c c c c||} 
 
 \hline
 Spaxel & $\rm V_1 \pm \Delta V_1~(\rm km~s^{-1})$ & $\rm V_2 \pm \Delta V_2~(\rm km~s^{-1})$ & $\rm \sigma_1 \pm \Delta\sigma_1~(\rm km~s^{-1})$ & $\rm \sigma_2 \pm \Delta\sigma_2~(\rm km~s^{-1})$ \\ [0.5ex] 
 \hline\hline
 Spaxel 1 (H$\alpha$)  & -2.7671$\pm$1.14 & -2.0153$\pm$5.28 & 3.603$\pm$5.40 & 3.878$\pm$ 1.14 \\
 \hline
 Spaxel 1 (NII) & 238.34$\pm$1.33 & 208.225$\pm$9.18 & 44.54$\pm$1.05 & 74.47$\pm$1.35 \\ 
  \hline
 
 \hline\hline
 Spaxel 2 (H$\alpha$) & -220.52$\pm$1.72 & -113.65$\pm$15.92 & 34.95$\pm$4.19 & 90.50$\pm$1.72 \\ 
 \hline
 Spaxel 2 (NII) & -242.04$\pm$5.73 & -170.61$\pm$10.42 & 48.34$\pm$1.53 & 38.85$\pm$5.73 \\ 
 \hline
 
 \hline\hline
 Spaxel 3 (H$\alpha$) & -248.59$\pm$1.50 & -140.43$\pm$7.74 & 66.76$\pm$15.01 & 69.92$\pm$1.50 \\
 \hline
 Spaxel 3 (NII) & -252.36$\pm$1.98 & -175.73$\pm$12.18 & 35.82$\pm$5.18 & 47.31$\pm$1.98 \\ 
 \hline
 
 \hline\hline
 Spaxel 4 (H$\alpha$)  & -278.44$\pm$3.16 & -196.77$\pm$6.57 & 200.93$\pm$4.51 & 73.13$\pm$3.16 \\ 
 \hline
 Spaxel 4 (NII) & -267.96$\pm$4.07 & -176.50$\pm$2.13 & 82.67$\pm$4.51 & 58.05$\pm$1.64 \\ 
 \hline
 
 \hline\hline
 Spaxel 5 (H$\alpha$)  & -291.78$\pm$2.10 & ---- & 122.62 $\pm$4.62 & ---- \\
 \hline
 Spaxel 5 (NII) & -251.27$\pm$0.98 & -58.79 $\pm$ 6.48 & 5.085 $\pm$ 4.07 & 1.3221 $\pm$0.98 \\
 \hline
 
 \hline\hline
 Spaxel 6 (H$\alpha$) & -73.24$\pm$4.35 & 88.08$\pm$2.37 & 193.27$\pm$8.75 & 276.89$\pm$4.35 \\ 
 \hline
 Spaxel 6 (NII) & -263.89 $\pm$9.87 & -73.19$\pm$1.17 & 78.04$\pm$8.41 & 118.16$\pm$9.87 \\ 
 \hline
 
 \hline\hline
 Spaxel 7 (H$\alpha$) & 132.66$\pm$7.47 & 313.36$\pm$5.72 & 93.72$\pm$6.07 & 46.42$\pm$7.47 \\
 \hline
 Spaxel 7 (NII) & -14.07$\pm$1.67 & 22.35$\pm$3.09 & 202.62$\pm$2.47 & 274.48$\pm$16.43 \\ 
 \hline
 
 \hline\hline
 Spaxel 8 (H$\alpha$) & 228.31$\pm$14.59 & 120.40$\pm$5.19 & 57.39$\pm$1.03 & 73.28$\pm$1.45 \\
 \hline
 Spaxel 8 (NII) & 300.75$\pm$6.15 & 117.76$\pm$10.83 & 66.50$\pm$6.06 & 136.84$\pm$6.15 \\ 
 \hline
 
 \hline\hline
 Spaxel 9 (H$\alpha$) &  277.58$\pm$19.22 & 113.62$\pm$2.13 & 84.19$\pm$1.35 & 72.06$\pm$1.92 \\
 \hline
 Spaxel 9 (NII) & 287.129$\pm$4.40 & 130.03$\pm$8.94 & 32.69$\pm$2.67 & 66.05$\pm$4.40 \\ 
  \hline
 
 \hline\hline
\end{tabular}
\end{center}
\end{table*}

\subsection {Rotation curves and spatially resolved velocity maps from disk model }  \label{appen:velocity}
 
\begin{figure}
	\includegraphics[width=\columnwidth]{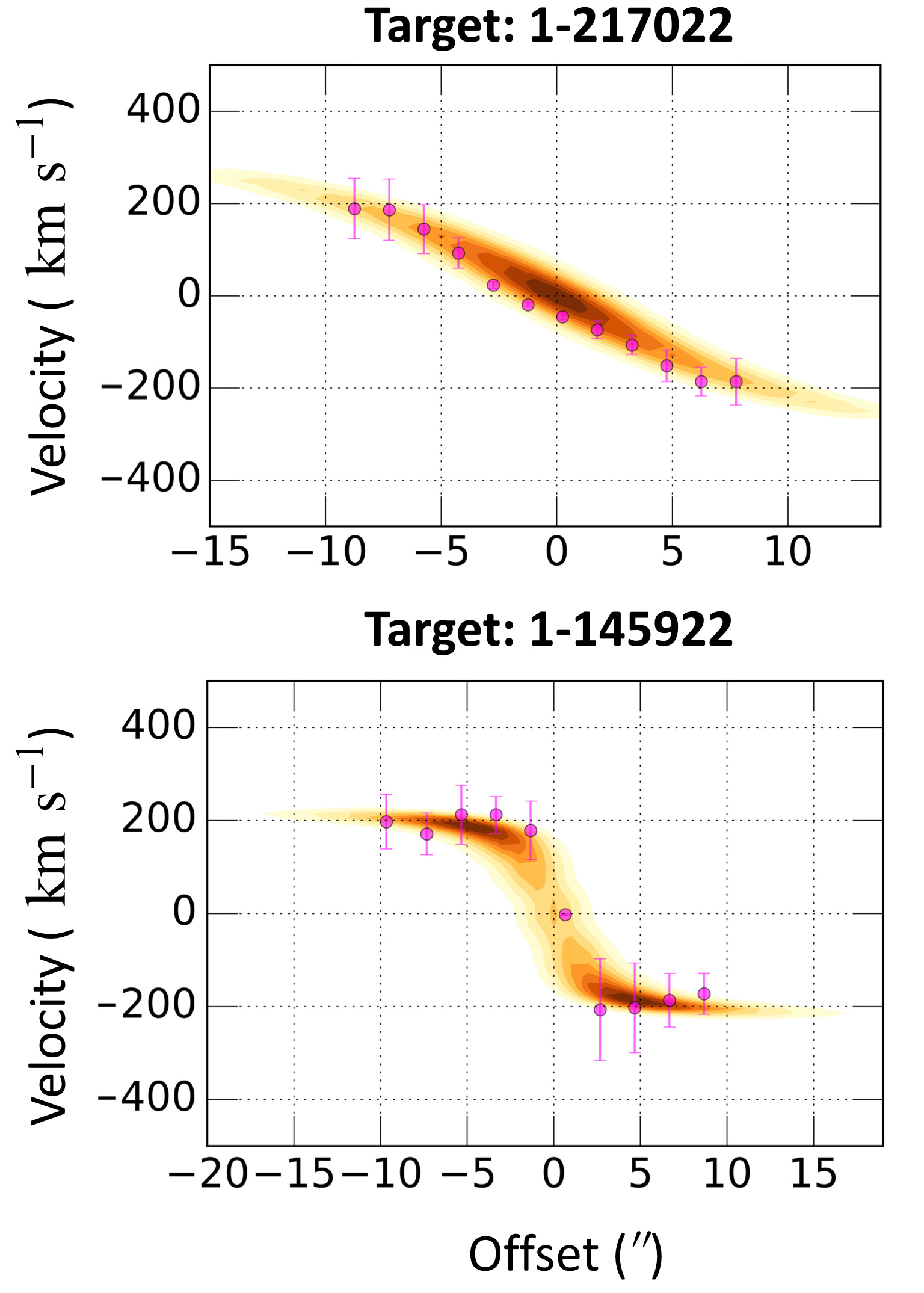}
    \caption{The position-velocity curve from the ESI data in magenta circles over-plotted on the best matched rotation curve assuming a disk model (details in \S \ref{sec:discussion}) in yellow contours for both target galaxies. The parameters of the disk model are chosen based on this qualitative match.  }
    \label{fig:pv}
\end{figure}

\begin{figure}
	\includegraphics[width=\columnwidth]{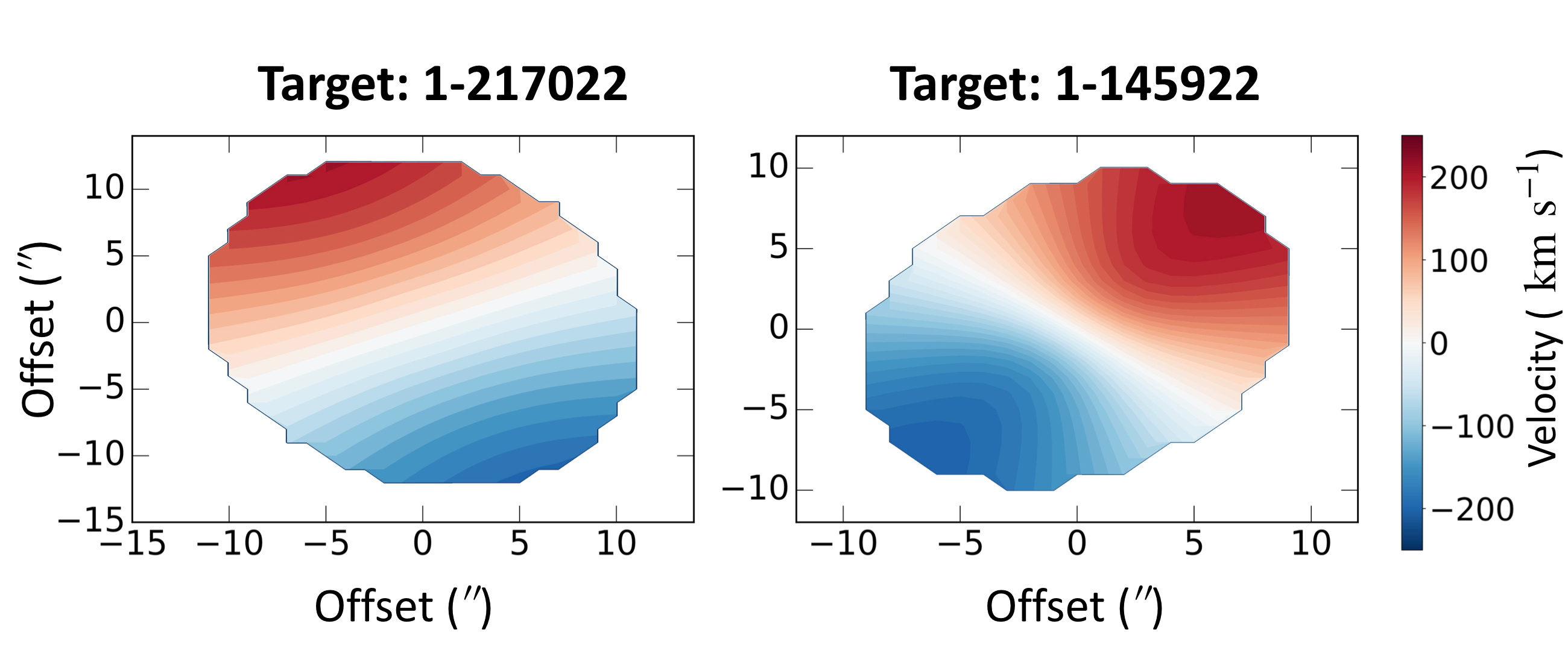}
    \caption{The spatially resolved 2D gas velocity map obtained from the best matched disk model for the first (left panel) and second (right panel) target galaxies.  }
    \label{fig:disk2D}
\end{figure}

Using the KINMSpy package, we construct a thin disk model defined by turnover radius, thickness of the disk, maximum velocity attained by the rotation curve and the scale radius associated with the assumed surface brightness profile for each of our target galaxies. The set of parameters chosen to best represent the observed target galaxies (see \S \ref{sec:discussion} for details) are decided based on the qualitative match between the modelled rotation curve and the observed velocity curve from ESI. The comparison between the ESI data (magenta circles) and that obtained from the model (yellow contours) is shown in Fig.~\ref{fig:pv}. The fair match between the rotation curves confirm the best possible choice of the disk parameters in an attempt to reproduce the data. Fig.~\ref{fig:disk2D} shows the corresponding 2D spatially resolved velocity map of the modelled gas particles obtained from the thin disk model using the chosen set of parameters. The observed 2D velocity fields are harder to reproduce using disk model for both galaxies.

\end{document}